\documentclass[aip, amsmath,amssymb, reprint, author-year]{revtex4-1}

\usepackage{graphicx}
\usepackage{dcolumn}
\usepackage{bm}

\def\kms{\hbox{km$\;$s$^{-1}$}}

\newcommand{\mss}{\scriptscriptstyle\mathsc}
\DeclareMathAlphabet{\mathsc}{OT1}{cmbr}{n}{sc}

\begin{document}


\title[Solar partially ionized plasma]{Fluid description of multi-component solar partially ionized plasma}

\author{E. Khomenko}
\email{khomenko@iac.es}
\author{M. Collados}
\affiliation{Instituto de Astrof\'{\i}sica de Canarias, 38205 La Laguna, Tenerife, Spain}
\affiliation{Departamento de Astrof\'{\i}sica, Universidad de La Laguna, 38205, La Laguna, Tenerife, Spain}
\author{A. D\'{\i}az}
\affiliation{Departament de F\'{\i}sica, Universitat de les Illes Balears, E-07122 Palma de Mallorca, Spain}
\author{N. Vitas}
\affiliation{Instituto de Astrof\'{\i}sica de Canarias, 38205 La Laguna, Tenerife, Spain}
\affiliation{Departamento de Astrof\'{\i}sica, Universidad de La Laguna, 38205, La Laguna, Tenerife, Spain}

\date{\today}

\begin{abstract}
We derive self-consistent formalism for the description of multi-component partially ionized solar plasma, by means of the coupled equations for the charged and neutral components for an arbitrary number of chemical species, and the radiation field. All approximations and assumptions are carefully considered. Generalized Ohm's law is derived for the single-fluid and two-fluid formalism.  Our approach is analytical with some order-of-magnitude support calculations. After general equations are developed we particularize to some frequently considered cases as for the interaction of matter and radiation.
\end{abstract}



 \maketitle

\section{Introduction}

The plasma in the lower atmosphere of the Sun (photosphere and chromosphere) is only partially ionized. The presence of neutrals is extremely high in the photosphere, where the estimations based on the standard atmospheric models provide about one ion every ten thousand neutrals \citep[see, e.g.][]{valc}. Higher up in the chromosphere the ionization fraction increases, but it remains below unity up to the transition region and corona. This fact alone does not necessarily imply that the magnetohydrodynamic (MHD) approximation (with some modifications) can not be safely applied. There are numerous examples of the success of MHD models describing basic processes in the photosphere and below, such as convection, magneto-convection, formation of magnetic structures, flux emergence, wave propagation, etc. \citep{Asplund+etal2000, Khomenko+Collados2006, Cheung+etal2007, Moreno-Insertis+etal2008, Nordlund+etal2009}. Such description is possible because, despite the presence of a significant amount of neutrals, the plasma is strongly collisionally coupled in the photosphere and below. However, when dealing with strongly magnetized regions, such as sunspots, one has to be careful  since the cyclotron frequencies of electrons and ions may become large enough to overcome the collisional frequencies leading to a break of assumptions of ideal MHD even in the photosphere.

The situation is different in the chromosphere because the collisional coupling of the plasma weakens with height. The physics of this region is less understood and the models of chromospheric phenomena are not so well developed \citep{Carlsson2007, Leenaarts2010}. In the chromosphere, the plasma changes from pressure- to magnetic field-dominated and, due to the weakening of collisional coupling, the electrons and ions become magnetized starting from some height, even for regions with relatively weak magnetic field. All these factors together mean that the assumptions of MHD and the approach based on the thermodynamic equilibrium conditions become invalid in the chromosphere.

It has long been known that the chromospheric plasma can not be treated in the local thermodynamic equilibrium (LTE) approach for processes related to the transfer of radiative energy and interaction of plasma and radiation. In recent years it has also become accepted that, for the description of the plasma processes themselves, the MHD approach, based on the total collisional coupling of the plasma, fails and alternative models must be applied. On the extreme edge of modelling, there is the kinetic description of the plasma. This, however, can not be of practical use in the photosphere and the chromosphere because these regions are far too dense. The intermediate approach consists in using the fluid-like description. Different modifications of such description include single-fluid quasi-MHD equations, or multi-fluid equations. In the multi-fluid approach, the equations for different plasma components - ions, neutrals and electrons - are solved separately, coupled by means of the collisional terms. In the quasi-MHD approach, the equations for different components are added up together leading to appearance of dissipative terms in the induction and energy conservation equations. In both cases, these equations need to be coupled to the equations describing the radiation field.

There is an increasing number of works in the literature where multi-fluid or single-fluid approaches have been applied for the description of various phenomena in the solar atmosphere. Deviations from classical MHD are found to be important for: (i) propagation of different types of high-frequency waves {\citep{Kumar+Roberts2003, Khodachenko2004, Khodachenko2006, Forteza2007, Pandey2008, Vranjes2008, Soler2009, Soler2010, Zaqarashvili2011, Zaqarashvili+etal2012}, as the relative motion between the neutral and charged species increases the collisional damping of these waves in the photosphere, chromosphere and prominence plasmas, modifies their excitation rates and produces cut-off frequencies; (ii) reconnection processes \citep{Zweibel1989, Brandenburg+Zweibel1994, Brandenburg+Zweibel1995, Leake+etal2012} since the collisional damping due to neutrals modifies the reconnection rates; reconnection studies in the two-fluid approach show that ions and neutrals can have rather different velocities and temperatures \citep{Sakai2006, Smith+Sakai2008, Sakai+Smith2009}; (iii) equilibrium balance of magnetic structures by facilitating the creation of potential force-free structures in the chromosphere \citep{Arber2009}, and offering new mechanisms for creation of intense photospheric flux tubes \citep{Khodachenko+Zaitsev2002}; (iv) magnetic flux emergence \citep{Leake+Arber2006, Arber2007} by increasing the amount of emerged flux due to the presence of a diffusive layer of partially ionized plasma in the photosphere; (v) plasma instabilities in prominences \citep{Soler+etal2012, Diaz+etal2012, Diaz+etal2013, Khomenko+etal2013, Khomenko+etal2014} by removing the critical wavelength and making the plasma unstable at all spatial scales; (vi) heating of chromospheric plasma \citep{DePontieu1998, Judge2008, Krasnoselskikh2010, Khomenko+Collados2012, MartinezSykora+etal2012} since the Joule dissipation of electric currents is enhanced by orders of magnitude due to the presence of neutrals and weakening of the collisional coupling.

\parskip 0pt

Because this subject is gaining so much attention, it is important to revise the derivation of the basic equations for the description of the multi-component solar plasma using a common frame of assumptions and simplifications. The purpose of the present article is to provide such derivation. With no criticism to any of the above mentioned studies, the equations are often taken from different sources, leading to a different formulation of various terms.   \citet{Meier+Shumlak2012} discussed recently the same problem and provided a formalism for the description of multi-component solar plasma, including ionization and recombination processes, aiming at general plasma physics applications. In our approach, the macroscopic equations of motion for the plasma components are self-consistently derived starting from Boltzmann equation. We include the interaction with radiation from the very beginning, via excitation/deexcitation and ionization/recombination processes and treat photons and yet another type of particles interacting with the rest of the mixture. The generalized Ohm's law is derived for the single-fluid and two-fluid cases. Unlike previous works, we consider not only hydrogen (or hydrogen-helium) plasma, but generalize to the case of multiple species in different ionization and excitation states. We then make order of magnitude estimates of the different terms in the generalized Ohm's law for the typical solar conditions.

\section{Macroscopic equations}

We consider a plasma composed of a mixture of atoms of different atomic species, neglecting molecules. These atoms can be excited to excitation levels and/or to different ionization stages. Each particle is determined by three numbers: its atomic species $\alpha$, its ionization stage $\mathsc{I}$ and its excitation level $\mathsc{E}$, defining its "micro-state", $\{ \alpha \mathsc{IE} \}$\footnote{ Through the paper we will use $\mathsc{I}=0$ for neutral atoms, $\mathsc{I}=1$ for singly-ionized ions, etc., so that ionization stage $\mathsc{I}$ gives $\mathsc{I}$ electrons. Similarly, we use $\mathsc{E}=0$ to indicate the ground state, and $\mathsc{E}>0$ for subsequent excitation states.}. The case of electrons and photons will be considered apart. The following relation is applied:
\begin{equation}
n_{\alpha}=\sum_{I}n_{\alpha \mss{I}}=\sum_{I,E}n_{\alpha \mss{IE}}
\end{equation}
where $n_{\alpha \mss{IE}}$ is the number density of particles of element $\alpha$ with ionization stage $\mathsc{I}$ and excitation level $\mathsc{E}$; $n_{\alpha \mathsc{I}}$ is the number density of particles of element $\alpha$ and ionization stage $\mathsc{I}$, and $n_{\alpha}$ is the number density of particles of element $\alpha$ in all ionization and excitation stages. 
The number of electrons is equal to the number of ions of all species and ionization states, taking into account that the $\mathsc{I}$'s ionization state gives $\mathsc{I}$ electrons: 
\begin{equation}
n_e=\sum_{\alpha,I}I\cdot n_{\alpha \mss{I}}
\end{equation}

\noindent The general transport equation of a scalar quantity $\chi(\vec{v})$ is derived from the Boltzmann equation, written for an ensemble of particles of a given micro-state $\{ \alpha \mathsc{IE} \}$:
\begin{equation} \label{eq:boltzman}
\frac{\partial f_{\alpha \mss{IE}}}{\partial t} + \vec{v} \vec{\nabla} f_{\alpha \mss{IE}} + \vec{a} \vec{\nabla}_v f_{\alpha \mss{IE}} = \left(\frac{\partial f_{\alpha \mss{IE}}}{\partial t}\right)_{\rm coll}
\end{equation}
\noindent where $f_{\alpha \mss{IE}}(\vec{r},\vec{v}, \vec{t})$  is the one-particle distribution function, $\vec{a}$ is acceleration and $\vec{v}$ is the particle velocity. The term on the right hand side of Eq.~\ref{eq:boltzman} describes the variation of the distribution function of our ensemble of particles due to collisions with other particles, not belonging to $\{ \alpha \mathsc{IE} \}$.
The collisional term must satisfy the usual conditions of the conservation of particles, their momentum and energy, see e.g. the textbooks by \citet{Goedbloed+Poedts2004} and \citet{Balescu}. 

The general transport equation for a scalar quantity $\chi(\vec{v})$ can be obtained by multiplying the Boltzmann equation by $\chi(\vec{v})$ and integrating it over the whole velocity space \citep{Braginskii1965, Bittencourt, Balescu}. We assume that the variable $\chi(\vec{v})$ does not explicitly depend on time nor space variables ($\partial \chi / \partial t=0$, $\vec{\nabla}\chi=0$). 
The transport equation then takes the form:
\begin{eqnarray} \label{eq:transport}
\frac{\partial}{\partial t}(n_{\alpha \mss{IE}}\langle\chi\rangle_{\alpha \mss{IE}}) + \vec{\nabla} (n_{\alpha \mss{IE}}\langle\chi\vec{v}\rangle_{\alpha \mss{IE}}) - n_{\alpha \mss{IE}}\langle\vec{a} \vec{\nabla}_v\chi\rangle_{\alpha \mss{IE}} \nonumber \\
=  \int_{V}{\chi \left(\frac{\partial f_{\alpha \mss{IE}}}{\partial t}\right)_{\rm coll} d^3 v}
\end{eqnarray}
\noindent Here, the triangular brackets, $\langle\rangle$, mean averaging of a quantity over the distribution function in velocity space.  Similar equation can be written for a vector quantity $\vec{\chi}(\vec{v})$:
\begin{eqnarray} \label{eq:transport_vector}
\frac{\partial}{\partial t} (n_{\alpha \mss{IE}} \langle \vec{\chi} \rangle_{\alpha \mss{IE}}) & +& \vec{\nabla}  (n_{\alpha \mss{IE}} \langle\vec{\chi}\otimes  \vec{v} \rangle_{\alpha \mss{IE}}) - n_{\alpha \mss{IE}} \langle (\vec{a} \vec{\nabla}_v) \vec{\chi} \rangle_{\alpha \mss{IE}} \nonumber \\
& = &\int_{V}{ \vec{\chi}\left(\frac{\partial f_{\alpha \mss{IE}}}{\partial t}\right)_{\rm coll} d^3 v}
\end{eqnarray}
\noindent where $\otimes$ represents the external (or tensorial) vector product. These two equations will be used below to derive the equations of conservations of mass, momentum and energy for particles in the micro-state $\{ \alpha \mathsc{IE} \}$.

Particle density, momentum and energy are defined as usual as the moments of the distribution function and their corresponding conservation equations are derived as the moments of the Boltzmann equation, see below. The average velocity $\vec{u}_{\alpha \mss{IE}}$ of each species in a given microstate $\{ \alpha \mathsc{IE} \}$ is obtained from the momentum, dividing by the corresponding particle's mass. 
The mass of particles is exactly the same for all excitation states $\mathsc{E}$, and differs between the ionization states by the mass of corresponding electrons. This gives only a small difference between the masses of particles in different ionization stages, but we will preserve it here for consistency.  We will assume that the macroscopic velocities of particles of a given excitation state are the same as of their corresponding ground state (either ionized or neutral) since there is no reason to expect that particles in excited state experience different forces than particle in the ground state. However, macroscopic velocities of the neutral and ionized states can be different, as the latter experience the action of the Lorentz force and the former do not:
\begin{eqnarray} \label{eq:mu-states}
m_{\alpha \mss{IE}} &=&m_{\alpha \mss{I}} \\ \nonumber
\vec{u}_{\alpha \mss{IE}}&=&\vec{u}_{\alpha \mss{I}} \neq \vec{u}_{\alpha \mss{I'}}\ \ {\rm (if}\ \ \mathsc{I} \ne \mathsc{I'})
\end{eqnarray}
It is also useful to decompose the velocity of a particle as the sum of the macroscopic $\vec{u}$ and random velocities $\vec{c}$:
\begin{equation} \label{eq:crandom}
\vec{v}=\vec{u}_{\alpha \mss{IE}}+\vec{c}_{\alpha \mss{IE}}=\vec{u}_{\alpha \mss{I}}+\vec{c}_{\alpha \mss{I}}
\end{equation}
Photons can be considered another type of particles and, as such, a similar Boltzmann equation can be written also for their distribution function $f_R$ \citep{Mihalas}:
\begin{equation} \label{eq:boltzman_R}
\frac{\partial f_R}{\partial t} + \vec{v} \vec{\nabla} f_R + \vec{\mathcal{F}} \vec{\nabla}_p f_R = \left(\frac{\partial f_R}{\partial t}\right)_{\rm coll}
\end{equation}
where the function $f_R=f_R(\vec{r},\vec{p},t)$ gives the number density of photons at the location $(\vec{r},\vec{r}+d\vec{r})$ and with momentum $(\vec{p},\vec{p}+d\vec{p})$. Equivalently, one can write $f_R=f_R(\vec{r},\vec{n},\nu,t)$ to represent the number density of photons at $(\vec{r},\vec{r}+d\vec{r})$, with frequency $(\nu,\nu+d\nu)$ propagating with velocity $c$ in the direction $\vec{n}$.

Since photons do not have mass at rest, if no relativistic effects are present, no force is acting on them, $\vec{\mathcal{F}}=0$, and photon propagation will be along straight lines with $\vec{v}=c\vec{n}$. Since each photon has energy $h\nu$ (being $h$ Planck constant), the amount of energy crossing a surface $dS$ in the direction $\vec{n}$ in the interval $dt$ is
\begin{equation}
dE= (c h \nu)f_R dS \cos(\theta) d\Omega d\nu dt
\end{equation}
where $\theta$ is the angle between the direction of $\vec{n}$ and the normal to the surface $dS$, and $\Omega$ is the solid angle. This gives us the relation between the photon distribution function and the specific intensity:
\begin{equation} \label{eq:intensity}
I_\nu(\vec{r},\vec{n},t)= c h \nu f_R(\vec{r},\vec{n},\nu,t)
\end{equation}
Substituting this relation into the Boltzmann equation (Eq.~\ref{eq:boltzman_R}) one obtains the usual form of the radiative transfer equation:
\begin{equation} \label{eq:rte}
\frac{1}{c h \nu}\left[\frac{\partial I_\nu}{\partial t} + c \vec{n} \vec{\nabla}I_\nu \right]= \left(\frac{\partial f_R}{\partial t}\right)_{\rm coll}
\end{equation}

The right-hand side of the equation is related to the generation/removal of photons in the radiation field due to excitation/de-excitation and ionization/recombination processes (see Eqs.~\ref{F_downward} and \ref{F_upward}). This collisional term is usually expressed as the difference between photon sources ($j_\nu$) and losses (which are proportional to the radiation field and are written as $k_\nu I_\nu$). The latter notation is the standard one in the literature on radiative transfer \citep[e.g.,][]{Mihalas}, and usually reads as
\begin{eqnarray}
\left(\frac{\partial f_R}{\partial t}\right)_{\rm coll}  = \frac{j_\nu - k_\nu I_\nu}{h \nu} 
\end{eqnarray}
After neglecting the temporal derivative of the radiation field in Eq. \ref{eq:rte} one gets the usual transfer equation:
\begin{equation} \label{eq:rte_standard}
\frac{d\,I_\nu}{d\,s} = j_\nu - k_\nu I_\nu
\end{equation}

\subsection{Mass conservation}

The equation of mass conservation for particles in a micro-state $\{ \alpha \mathsc{IE} \}$ is derived from Eq.~\ref{eq:transport} by setting $\chi=m_{\alpha \mss{IE}}=m_{\alpha \mss{I}}$. This way $\langle\chi\rangle_{\alpha \mss{IE}}=m_{\alpha \mss{I}}$, $\langle\chi\vec{v}\rangle_{\alpha \mss{IE}}=m_{\alpha \mss{I}}\langle\vec{v}_{\alpha \mss{I}}\rangle= m_{\alpha \mss{I}}\vec{u}_{\alpha \mss{I}}$ and $\vec{\nabla}_v\chi=0$ and one obtains:
\begin{equation} \label{eq:continuity-aie}
\frac{\partial \rho_{\alpha \mss{IE}}}{\partial t} + \vec{\nabla} (\rho_{\alpha \mss{IE}}\vec{u}_{\alpha \mss{I}}) = m_{\alpha \mss{I}}\int_{V}{\left(\frac{\partial f_{\alpha \mss{IE}}}{\partial t}\right)_{\rm coll} d^3 v} = S_{\alpha \mss{IE}}
\end{equation}
where $\rho_{\alpha \mss{IE}}=m_{\alpha \mss{I}}n_{\alpha \mss{IE}}$ and Eqs.~\ref{eq:mu-states} and \ref{eq:crandom} are taken into account for velocities.

The collision term on the right hand side ($S_{\alpha \mss{IE}}$) accounts for collisions with particles of another kind that lead to creation or destruction of particles of the micro-state $\{ \alpha \mathsc{IE} \}$, including the interaction with photons (i.e. radiation field) as another type of particles. If particle identity at the micro-state $\{ \alpha \mathsc{IE} \}$ is maintained during the collision, such collision is called ``elastic''. The term $S_{\alpha \mss{IE}}$ is zero for elastic collisions. The collisions that lead to creation/destruction of particles are called ``inelastic''. Inelastic processes most relevant for the solar atmosphere are ionization, recombination, excitation and de-excitation. Scattering can be treated as a particular case of photon absorption and re-emission by any of the above processes. Thus, in a general case, the collisional term can be written in the following form:
\begin{equation}
S_{\alpha \mss{IE}}= m_{\alpha \mss{I}}\sum_{\mss{I',E'}}(n_{\alpha \mss{I'E'}}P_{\alpha \mss{I'E' IE}} - n_{\alpha \mss{IE} }P_{\alpha \mss{IE I'E'}})
\end{equation}
where the summation goes over all ionization states $\mathsc{I'}$ and their corresponding excitation states $\mathsc{E'(I')}$ leading to creation/destruction of particles of micro-state $\{ \alpha \mathsc{IE} \}$ of the $\alpha$ element. The first term describes transitions from other excitation/ionization states of element $\alpha$ to the state $\mathsc{IE}$, and the second term describes the opposite processes. The rate coefficients $P_{\alpha \mss{IE I'E'}}$ are the sum of collisional ($C_{\alpha \mss{IE I'E'}}$) and radiative ($F_{\alpha \mss{IE I'E'}}$) rate coefficients:
\begin{equation}
P_{\alpha \mss{IE I'E'}}=F_{\alpha \mss{IE I'E'}}+C_{\alpha \mss{IE I'E'}}
\end{equation}

The radiative rates $F_{\alpha \mss{IE I'E'}}$ have to be specified separately for the upward (initial microstate with a lower energy than the final one) and for the downward processes. The former ones are photoexcitation ($\mathsc{E < E'}$) and photoionization ($\mathsc{I < I'}$), while the latter ones are photodeexcitation ($\mathsc{E > E'}$) and photorecombination  ($\mathsc{I > I'}$). The unified radiative rates $F_{\alpha \mss{IE I'E'}}$  can be found elsewhere in standard radiative transfer tutorials, e.g. \citet{Rutten2003,  Carlsson1986}. They depend on the radiation field and on the atomic parameters of the transitions.

The unified radiative rates $F_{\alpha \mss{IE I'E'}}$ for the upward  processes (excitation, ionization, $\mathsc{IE < I'E'}$) can be written as:
\begin{eqnarray}
F_{\alpha \mss{IE I'E'}} = \int_{0}^{\infty}\oint\frac{\sigma_{\alpha \mss{IE I'E'}}}{h\nu}  I_{\nu}d\Omega d\nu 
\label{F_downward}
\end{eqnarray}
The coefficient $\sigma_{\alpha \mss{IE I'E'}}$ is the extinction coefficient of interaction of an atom with a photon to give rise to the transition from state $\mathsc{IE}$ to $\mathsc{I'E'}$. The integral over angle and frequency comes the fact that photons coming from different directions or frequencies may be absorbed to produce a give rise to an excitation or ionization process. 
 
For the downward radiative processes (deexcitation, recombination,  $\mathsc{IE > I'E'}$) the unified rates $F_{\alpha \mss{IE I'E'}}$ are:
\begin{eqnarray}
F_{\alpha \mss{IE I'E'}} = \int_{0}^{\infty}\oint\frac{\sigma_{\alpha \mss{IE I'E'}} }{h\nu}  G_{\alpha \mss{IE I'E'}} \left(\frac{2h\nu^3}{c^2} + I_{\nu}\right)  d\Omega d\nu  \nonumber \\
\label{F_upward}
\end{eqnarray}

It is important to note that these expressions represent the important link between the radiation field and the atomic excitation/deexcitation/ionization/recombination processes. Through these radiative processes, the photon absorption/emission has to be taken into account in addition to the corresponding change in the population of the $\{\alpha \mathrm{IE}\}$ microstate.

The extinction coefficients $\sigma_{\alpha \mss{IE I'E'}}$  and the coefficient $G_{\alpha \mss{I'E' IE}}$ have different forms for bound-bound and bound-free processes and can be found in standard tutorials on radiative transfer. 
The rates for the inelastic collisions between a particle in microstate $\{\alpha \mathrm{IE}\}$ and another particle (collider $\beta$) that turn $\{\alpha \mathrm{IE}\}$ into $\{\alpha \mathrm{I'E'}\}$ may be  written in a general form as: 
\begin{equation}
C_{\alpha \mss{IE I'E'},\beta} =  n_\beta \int_{v_0}^\infty \sigma_{\alpha \mss{IE I'E'}}(v_\beta) f(v_\beta) v_\beta d v_\beta,
\end{equation}
\noindent where $f(v_\beta)$ is the velocity distribution of the colliders and $\sigma_{\alpha \mss{IE I'E'},\beta}$ is the collisional cross-section of species $\alpha$ and $\beta$. the total collisional rate coefficient, $\sigma_{\alpha \mss{IE I'E'}}$, is obtained after adding up the contribution off all colliders, i.e.,
\begin{equation}
C_{\alpha \mss{IE I'E'}} = \sum_\beta C_{\alpha \mss{IE I'E'},\beta} 
\end{equation}

The most common colliders in the solar atmosphere are the electrons. The collisional rates for the bound-bound and bound-free processes including free electrons are usually specified by the  approximation of \citet{VanRegemorter1962} \citep[see][]{Rutten2003},  derived assuming local thermodynamical equilibrium.

If we sum Eq.~\ref{eq:continuity-aie} over all possible microstates $\mathsc{IE}$ of the $\alpha$ atom, the right hand side becomes identical to zero as no nuclear reactions are considered and no transformation between the atoms is allowed ($\sum_{\mss{IE}}S_{\alpha \mss{IE}}=S_\alpha=0$). However, the right hand side of Eq.~\ref{eq:continuity-aie}  written for a specific microstate $S_{\alpha \mss{IE}}$ may also become zero under certain approximations. It is certainly identical to zero if the plasma is in thermodynamical equilibrium (TE) since in that case all processes are in detailed balance, so $n_{\alpha \mss{I'E'}}P_{\alpha \mss{I'E' IE}} = n_{\alpha \mss{IE} }P_{\alpha \mss{IE I'E'}}$. The same applies for the approximation of local thermodynamical equilibrium (LTE) where TE is assumed to be valid locally. Furthermore, it is still identical to zero when the LTE approximation is relaxed to the approximation of the statistical equilibrium (SE) that is commonly solved within the instantaneous non-LTE radiative transfer problem. All these approximations require that the macroscopic changes of the atmosphere happen on temporal and spatial scales such that the atmosphere may be considered stationary (${\partial \rho_{\alpha \mss{IE}}}/{\partial t} = 0$)
and static ($\vec{\nabla} (\rho_{\alpha \mss{IE}}\vec{u}_{\alpha \mss{I}}) = 0$).

The condition of SE may formally be written as 
\begin{equation}
S_{\alpha \mss{IE}}= m_{\alpha \mss{I}}\sum_{\mss{I',E'}}(n_{\alpha \mss{I'E'}}P_{\alpha \mss{I'E' IE}} - n_{\alpha \mss{IE}}P_{\alpha \mss{IE I'E'}}) = 0
\end{equation}
In a general case, the net source terms $S_{\alpha \mss{IE}}$ cannot be neglected in the solar chromosphere when the SE fails to  describe the plasma conditions \citep{Carlsson+Stein2002, Leenaarts2007} due to the imbalance between the ionization and recombination rates.  

Eq.~\ref{eq:continuity-aie} can be summed over the excitation states $\mathrm{E}$ to give:
\begin{equation} \label{eq:continuity-ai}
\frac{\partial \rho_{\alpha \mss{I}}}{\partial t} + \vec{\nabla} (\rho_{\alpha \mss{I}}\vec{u}_{\alpha \mss{I}}) =S_{\alpha \mss{I}}\end{equation}
where we define
\begin{equation}
S_{\alpha \mss{I}}=m_{\alpha \mss{I}}\sum_{\mss{E}}\sum_{\mss{I'\neq I,E'}}(n_{\alpha \mss{I'E'}}P_{\alpha \mss{I'E' IE}} - n_{\alpha \mss{IE}}P_{\alpha \mss{IE I'E'}}) 
\end{equation}

\noindent In this summation the contributions from the transitions between different excitation states $\mathsc{E(I)}$ of the same ionization state $\mathsc{I}$ cancel out. The remaining contributions include transitions that change the ionization state (whatever the excitation state) and therefore generate or remove a free electron. 

In the case of electrons,  by substituting $\chi=m_e$ into equation Eq.~\ref{eq:transport} we obtain:
\begin{equation}\label{eq:continuity-e}
\frac{\partial \rho_e}{\partial t} + \vec{\nabla} (\rho_e\vec{u}_e) = m_e\int_{V}{\left(\frac{\partial f_e}{\partial t}\right)_{\rm coll} d^3 v} = S_e
\end{equation}
The collisional term for electrons takes into account all the processes leading to the appearance/disappearance of electrons (ionization and recombination) for all atomic species $\alpha$ of the system.
\begin{equation}
S_e = m_e\sum_\alpha\sum_{\mss{I,E}}\ \sum_{\mss{I'>I,E'}}(\mathsc{I'}-\mathsc{I})(n_{\alpha \mss{IE}}P_{\alpha \mss{IEI'E'}} - n_{\alpha \mss{I'E'}}P_{\alpha \mss{I'E'IE}})
\end{equation}
This term only becomes strictly null in the case of TE/LTE/SE.

Since photons do not have mass, there is no equivalent mass conservation equation for them.

\subsection{Momentum conservation}
\label{subsec:Momentum}

By setting $\vec{\chi}=m_{\alpha \mss{I}}\vec{v}$ in Eq.~\ref{eq:transport_vector}, the equation of momentum conservation for particles of a micro-state $\{ \alpha \mathsc{IE} \}$ is derived. According to Eq.~\ref{eq:crandom}, we have $\vec{v}=\vec{u}_{\alpha \mss{IE}}+\vec{c}_{\alpha \mss{IE}}=\vec{u}_{\alpha \mss{I}}+\vec{c}_{\alpha \mss{I}}$ and $m_{\alpha \mss{IE}}=m_{\alpha \mss{I}}$. Using the continuity equation for the micro-state $\{ \alpha \mathsc{IE} \}$, Eq.~\ref{eq:continuity-aie}, and defining the kinetic pressure tensor, ${\bf\hat{p}}_{\alpha \mss{IE}}$, according to Eq.~\ref{eq:app1-p} (see Appendix~\ref{app:pq}), the equation of momentum conservation is obtained:
\begin{eqnarray}
\rho_{\alpha \mss{IE}}\left[\frac{\partial \vec{u}_{\alpha \mss{I}}}{\partial t} + (\vec{u}_{\alpha \mss{I}} \vec{\nabla})\vec{u}_{\alpha \mss{I}} \right] &+& \vec{\nabla}  {\bf\hat{p}}_{\alpha \mss{IE}} - n_{\alpha \mss{IE}}\langle\vec{\mathcal{F}}\rangle_{\alpha \mss{IE}} =   \\
m_{\alpha \mss{I}}\int_V{\vec{v}\left(\frac{\partial f_{\alpha \mss{IE}}}{\partial t}\right)_{\rm coll}d^3 v} &-& m_{\alpha \mss{I}}\vec{u}_{\alpha \mss{I}}\int_V{\left(\frac{\partial f_{\alpha \mss{IE}}}{\partial t}\right)_{\rm coll} d^3 v} \nonumber
\end{eqnarray}

\noindent Introducing the total derivative and assuming that the average force $\vec{\mathcal{F}}$ has an electromagnetic and gravitational nature, the equation of motion is written as:
\begin{eqnarray} \label{eq:momentum-aie}
\rho_{\alpha \mss{IE}}\frac{D\vec{u}_{\alpha \mss{I}}}{Dt} &= & n_{\alpha \mss{IE}}q_{\alpha \mss{I}}(\vec{E} + \vec{u}_{\alpha \mss{I}}\times\vec{B}) + \rho_{\alpha \mss{IE}}\vec{g} - \vec{\nabla}  {\bf\hat{p}}_{\alpha \mss{IE}} \nonumber \\
&+& \vec{R}_{\alpha \mss{IE}} - \vec{u}_{\alpha \mss{I}}S_{\alpha \mss{IE}}.
\end{eqnarray}
where
\begin{equation}
\vec{R}_{\alpha \mss{IE}}=m_{\alpha \mss{I}}\int_V{\vec{v}\left(\frac{\partial f_{\alpha \mss{IE}}}{\partial t}\right)_{\rm coll}d^3 v}
\end{equation}

\noindent The term $\vec{R}_{\alpha \mss{IE}}$ provides the momentum exchange due to collisions of particles in the micro-state $\{\alpha \mathsc{IE} \}$ with other particles, including photons.  The expressions for elastic collisions between neutrals, ions and electrons (excluding photons) are given in Appendix~\ref{app:SR}.  

The above equation can be summed up for all excitation states $\mathsc{E}$. In this summation it is important to take into account that we assumed that macroscopic velocities and masses of all excitation states are the same. This makes the summation straightforward. The following momentum equation is then obtained:
\begin{eqnarray} \label{eq:momentum-ai}
\rho_{\alpha \mss{I}}\frac{D\vec{u}_{\alpha \mss{I}}}{Dt} &= &n_{\alpha \mss{I}}q_{\alpha \mss{I}}(\vec{E} + \vec{u}_{\alpha \mss{I}}\times\vec{B}) + \rho_{\alpha \mss{I}}\vec{g} - \vec{\nabla}  {\bf\hat{p}}_{\alpha \mss{I}} \nonumber \\
&+& \vec{R}_{\alpha \mss{I}} -\vec{u}_{\alpha \mss{I}}S_{\alpha \mss{I}}
\end{eqnarray}
where $\sum_E \vec{R}_{\alpha \mss{IE}}=\vec{R}_{\alpha \mss{I}}$ (Appendix~\ref{app:SR}), and other definitions are given in Appendix~\ref{app:pq}.

For electrons, an equation similar to Eq.~\ref{eq:momentum-aie} can be obtained introducing $\vec{\chi}=m_e\vec{v}$:
\begin{equation} \label{eq:momentum-e}
\rho_e\frac{D\vec{u}_e}{Dt}=-en_e(\vec{E} + \vec{u}_e\times\vec{B}) + \rho_e\vec{g} - \vec{\nabla}  {\bf\hat{p}}_e + \vec{R}_e - \vec{u}_eS_e
\end{equation}
where we have used $q_e=-e$.

To get the momentum conservation for photons, we can multiply Eq.~\ref{eq:rte} by $\vec{n}$, the propagation direction of photons, and integrate over all solid angles and all frequencies. With this, one gets the equation
\begin{equation}
\frac{1}{c^2}\frac{\partial \vec{F}_R}{\partial t} + \vec{\nabla}{\bf{\hat{P}}}_R =\frac{1}{c} \int_0^{\infty} \oint (j_\nu - k_\nu I_\nu)\vec{n} d\Omega d\nu
\end{equation}
\noindent  where the radiative energy flux, $\vec{F}_R$, is defined as
\begin{equation}
\vec{F}_R(\vec{r},t)=  \int_0^{\infty} \oint \vec{n}I_\nu(\vec{r},\vec{n},\nu,t)d\Omega d\nu
\end{equation}
\noindent and the radiation pressure, ${\bf{\hat{P}}}_R$, as
\begin{equation}
{\bf{\hat{P}}}_R(\vec{r},t)= \frac{1}{c} \int_0^{\infty} \oint \vec{n}\otimes \vec{n} I_\nu(\vec{r},\vec{n},\nu,t)d\Omega d\nu
\end{equation}
Neglecting the time variations of the radiation field, one gets the momentun conservation equation for photons,
\begin{eqnarray}
\label{eq:photon_momentum}
\vec{\nabla}{\bf{\hat{P}}}_R &=& 
\frac{1}{c} \int_0^{\infty} \oint (j_\nu - k_\nu I_\nu)\vec{n} d\Omega d\nu 
\end{eqnarray}

The relation between the collisional momentum term $\vec{R}_{\alpha \mss{IE}}$ and the radiation field depends on particular transition, and it is of no use here to introduce the complexity to describe all possible transitions, that can be found elsewhere.
For the purpose of illustration, we provide a simple example of a radiative excitation process, whereby an atom passes from micro-state $\{\alpha \mathsc{IE'} \}$ to microstate $\{\alpha \mathsc{IE} \}$ after the absorption of a photon. With this process, the momentum of microstate $\{\alpha \mathsc{IE} \}$ is increased with the initial momentum of the particles in micro-state $\{\alpha \mathsc{IE'} \}$ plus the momentum of the absorbed photon. For this particular example,  the contribution $\vec{R}_{\alpha \mss{IE}}^{\rm rad}$  to the term $\vec{R}_{\alpha \mss{IE}}$ by this process would be 
\begin{equation} \label{eq:r-photons-example}
\vec{R}_{\alpha \mss{IE}}^{\rm rad} = n_{\alpha \mss{IE'}} m_{\alpha \mss{I}} \vec{u}_{\alpha \mss{I}} F_{\alpha \mss{IE'IE}}
 + \frac{1}{c} \int_{0}^{\infty} \oint \sigma_{\alpha \mss{IE' IE}} I_{\nu} d\Omega d\nu
\end{equation}
The integral in solid angle takes into account that photons coming from different directions can be absorbed, while the integral in frequency includes all possible photons that may lead to the transition through the cross-section factor $\sigma_{\alpha \mss{IE' IE}}$. Similar expressions can be written for the other radiative processes leading to changes of the momentum of each microstate $\{\alpha \mathsc{IE} \}$. The total value for $\vec{R}_{\alpha \mss{IE}}$ is obtained after adding up the contributions of all processes that populate/depopulate microstate $\{\alpha \mathsc{IE} \}$.

\subsection{Energy conservation}

In a general case, the gas internal energy of a given micro-state makes up of kinetic energy of particle motion, potential energy of their interaction (excitation states) and ionization energy \citep{Mihalas+Mihalas}. For an ideal gas, the potential and ionization energies of the particles are neglected and the internal energy of a gas is only due to random thermal motions:
\begin{equation}
e_{\alpha \mss{IE}}=\rho_{\alpha \mss{IE}} \langle c_{\alpha \mss{I}}^2 \rangle/2
\end{equation}
Taking into account the definition of the scalar pressure, $p_{\alpha \mss{IE}}=\frac{1}{3} \rho_{\alpha \mss{IE}}\langle c_{\alpha
\mss{I}}^2\rangle$, the internal energy can be expressed as:
\begin{equation}
e_{\alpha \mss{IE}}=\frac{3}{2}p_{\alpha \mss{IE}}
\end{equation}

\noindent In a more general situation, the internal energy will make up of kinetic energy and potential energy of the excitation-ionization level $\mathsc{IE}$, $E_{\alpha \mss{IE}},$  with respect to the ground level of the neutral state, $E_{\alpha \mss{00}}$:
\begin{equation}
e_{\alpha \mss{IE}}=\frac{3}{2}p_{\alpha \mss{IE}} + n_{\alpha \mss{IE}} E_{\alpha \mss{IE}}
\label{eq:eint-aie-def}
\end{equation}

\noindent To derive the conservation law for the quantity given by Eq.~\ref{eq:eint-aie-def}, we here take:
\begin{equation}
\chi=\chi_1+\chi_2=m_{\alpha \mss{I}}v^2/2 + E_{\alpha \mss{IE}}
\end{equation}
The potential energy associated to a given microstate is a scalar constant.

We start by setting $\chi=\chi_1=m_{\alpha \mss{I}}v^2/2$ in Eq.~\ref{eq:transport}. Following the standard steps, the energy equation is written:
\begin{eqnarray} \label{eq:ene0}
\frac{\partial}{\partial t} (\frac{3}{2}p_{\alpha \mss{IE}} + \frac{1}{2} \rho_{\alpha \mss{IE}}u_{\alpha \mss{I}}^2) + \vec{\nabla}(\frac{1}{2}\rho_{\alpha \mss{IE}}\langle {v^2} \vec{v} \rangle_{\alpha \mss{IE}}) &-& n_{\alpha \mss{IE}} \langle\vec{\mathcal{F}}\vec{v}\rangle_{\alpha \mss{IE}} \nonumber \\
= \frac{1}{2} m_{\alpha \mss{I}} \int_V{v^2\left(\frac{\partial f_{\alpha \mss{IE}}}{\partial t}\right)_{\rm coll} d^3 v} 
\end{eqnarray}
where $\vec{\mathcal{F}}$ is a general force. We rewrite the second term on the left hand side using the definition of the mean and random velocities (Eq.~\ref{eq:crandom}), pressure tensor (${\bf\hat{p}}_{\alpha \mss{IE}}$, Eq.~\ref{eq:app1-p}) and heat flow vector ($\vec{q}_{\alpha \mss{IE}}$, Eq.~\ref{eq:app1-q}). Under the additional assumption that any external force is independent of velocity or does not have a component parallel to the velocity (which includes both, electromagnetic and gravitational, forces), combining the above equation with those of mass (Eq.~\ref{eq:continuity-aie}) and momentum conservation (Eq.~\ref{eq:momentum-aie}) gives:
\begin{eqnarray} \label{eq:energy-p}
\frac{D}{D t}\frac{3 p_{\alpha \mss{IE}}}{2} +\frac{3}{2} p_{\alpha \mss{IE}}\vec{\nabla} \vec{u}_{\alpha \mss{I}}+ ({\bf\hat{p}}_{\alpha \mss{IE}}\vec{\nabla})\vec{u}_{\alpha \mss{I}} + \vec{\nabla}\vec{q}_{\alpha \mss{IE}}= \nonumber \\
M_{\alpha \mss{IE}} - \vec{u}_{\alpha \mss{I}} \vec{R}_{\alpha \mss{IE}} + \frac{1}{2}u_{\alpha \mss{I}}^2 S_{\alpha \mss{IE}}
\end{eqnarray}
where
\begin{equation}
M_{\alpha \mss{IE}}=\frac{1}{2} m_{\alpha \mss{I}} \int_V{v^2\left(\frac{\partial f_{\alpha \mss{IE}}}{\partial t}\right)_{\rm coll} d^3 v} 
\end{equation}

This term is zero in \citet{Braginskii1965}. However, in our point of view, it should be retained, given that the term $M_{\alpha \mss{IE}}$ includes kinetic energy losses/gains of the particles of micro-state $\{\alpha  \mathsc{IE} \}$ due to collisions with other particles, which in general will not vanish.

Now we consider the second contribution to the internal energy and set $\chi=\chi_2=E_{\alpha \mss{IE}}$ in equation Eq.~\ref{eq:transport}. We obtain for a given micro-state:
\begin{equation} \label{eq:energy-ion}
\frac{D n_{\alpha \mss{IE}}E_{\alpha \mss{IE}}}{D t}+ n_{\alpha \mss{IE}}E_{\alpha \mss{IE}} \vec{\nabla}\vec{u}_{\alpha \mss{I}} =E_{\alpha \mss{IE}}S_{\alpha \mss{IE}}/m_{\alpha \mss{I}}
\end{equation}
The term $E_{\alpha \mss{IE}}S_{\alpha \mss{IE}}$ takes into account the change of potential energy of a micro-state due to inelastic collisions with other particles, including photons, i.e. potential energy change during radiative and collisional ionization, recombination, excitation and deexcitation. 

Adding up the energy equations Eqs.~\ref{eq:energy-p} and \ref{eq:energy-ion} and using the definition of the internal energy of a micro-state $\{\alpha \mathsc{IE}\}$ from Eq.~\ref{eq:eint-aie-def} we rewrite the conservation equation for internal energy like follows:
\begin{equation} \label{eq:energy-aie}
\frac{D e_{\alpha \mss{IE}}}{D t} + e_{\alpha \mss{IE}}\vec{\nabla}\vec{u}_{\alpha \mss{I}} + {\bf\hat{p}}_{\alpha \mss{IE}}\vec{\nabla}\vec{u}_{\alpha \mss{I}} + \vec{\nabla}\vec{q}_{\alpha \mss{IE}} = Q_{\alpha \mss{IE}}
\end{equation}

\noindent where we defined the internal energy losses/gains term as:
\begin{equation}
Q_{\alpha \mss{IE}}=M_{\alpha \mss{IE}} - \vec{u}_{\alpha \mss{I}}\vec{R}_{\alpha \mss{IE}} +  \left(\frac{1}{2} u_{\alpha \mss{I}}^2 + E_{\alpha \mss{IE}}/m_{\alpha \mss{I}}\right) S_{\alpha \mss{IE}}
\end{equation}

In this definition, the $Q_{\alpha \mss{IE}}$ term includes energy losses/gains due to: elastic collisions of particles of the micro-state $\{\alpha \mathsc{IE}\}$ with other particles ($M_{\alpha \mss{IE}}$ term, and part of the $\vec{u}_{\alpha \mss{I}}\vec{R}_{\alpha \mss{IE}}$ term, Eq. \ref{eq:r-alpha-i}), and inelastic collisions with photons and other particles $(u_{\alpha \mss{I}}^2/2 + E_{\alpha \mss{IE}}/m_{\alpha \mss{I}} ) S_{\alpha \mss{IE}}$, and the corresponding part of the $\vec{u}_{\alpha \mss{I}}\vec{R}_{\alpha \mss{IE}}$ term, see Eq. \ref{eq:r-photons-example} for an example.

Adding up Eq.~\ref{eq:energy-aie} for all excitation states one obtains:
\begin{eqnarray} \label{eq:energy-ai}
\frac{D e_{\alpha \mss{I}}}{D t} + e_{\alpha \mss{I}}\vec{\nabla}\vec{u}_{\alpha \mss{I}} + {\bf\hat{p}}_{\alpha \mss{I}}\vec{\nabla}\vec{u}_{\alpha \mss{I}} + \vec{\nabla}\vec{q}_{\alpha \mss{I}} = Q_{\alpha \mss{I}}
\end{eqnarray}
\noindent with
\begin{eqnarray}
Q_{\alpha \mss{I}}=M_{\alpha \mss{I}} - \vec{u}_{\alpha \mss{I}}\vec{R}_{\alpha \mss{I}}+\frac{1}{2} u_{\alpha \mss{I}}^2S_{\alpha \mss{I}} + \Phi_{\alpha \mss{I}}
\end{eqnarray}
and
\begin{equation}
e_{\alpha \mss{I}}=3p_{\alpha \mss{I}}/2 + \chi_{\alpha \mss{I}} 
\end{equation}
where $\chi_{\alpha \mss{I}}$ and  $\Phi_{\alpha \mss{I}}$  are defined according to Eq. \ref{eq:chi-ai} and Eq. \ref{eq:Phi-ai}, see Appendix \ref{app:pq}.

For electrons, the above equation simplifies even further since there are no ionization-excitation states, neither potential energy corresponding to such states.
\begin{eqnarray}
\frac{D e_e}{D t} + e_e\vec{\nabla}\vec{u}_e +  {\bf\hat{p}}_e\vec{\nabla}\vec{u}_e + \vec{\nabla}\vec{q}_e =& &\nonumber \\
M_e -  \vec{u}_e\vec{R}_e & +& \frac{1}{2}u_e^2 S_e
\label{eq:energy-e}
\end{eqnarray}
with $e_e=3p_e/2$. 

The energy equation for photons is obtained after integrating Eq.~\ref{eq:rte} for all solid angles and for all frequencies 
\begin{equation}
\frac{\partial E_R}{\partial t} + \vec{\nabla}\vec{F}_R = \int_0^{\infty} \oint (j_\nu - k_\nu I_\nu) d\Omega d\nu
\end{equation}
\noindent where the radiative energy is defined as
\begin{equation}
E_R(\vec{r},t)= \frac{1}{c} \int_0^{\infty} \oint I_\nu(\vec{r},\vec{n},\nu,t)d\Omega d\nu
\end{equation}
The standard equation is obtained after neglecting the temporal variation of the radiation field, to lead to the expression
\begin{equation}
\label{eq:photon_energy}
\vec{\nabla}\vec{F}_R = \int_0^{\infty} \oint (j_\nu - k_\nu I_\nu) d\Omega d\nu
\end{equation}
Again, we can reinforce the concept of the coupling between the plasma particles and the radiation field. To that aim, we can use the same example as in Sect. \ref{subsec:Momentum}, excitation processes whereby particles in microstate $\alpha \mathsc{IE'}$ change to microstate $\alpha \mathsc{IE}$ after absorbing a photon. In this case, we have
\begin{equation}
M_{\alpha \mss{IE}} = \frac{1}{2} n_{\alpha \mss{IE'}} m_{\alpha \mss{I}} {u}_{\alpha \mss{I}}^2 F_{\alpha \mss{IE'IE}}
\end{equation}
and
\begin{equation}
(E_{\alpha \mss{IE}} - E_{\alpha \mss{IE'}}) S_{\alpha \mss{IE}}/m_{\alpha \mss{I}} = \int_{0}^{\infty} \oint \sigma_{\alpha \mss{IE' IE}} I_{\nu} d\Omega d\nu
\end{equation}
%

\begin{figure}
\center
\includegraphics[width=9cm]{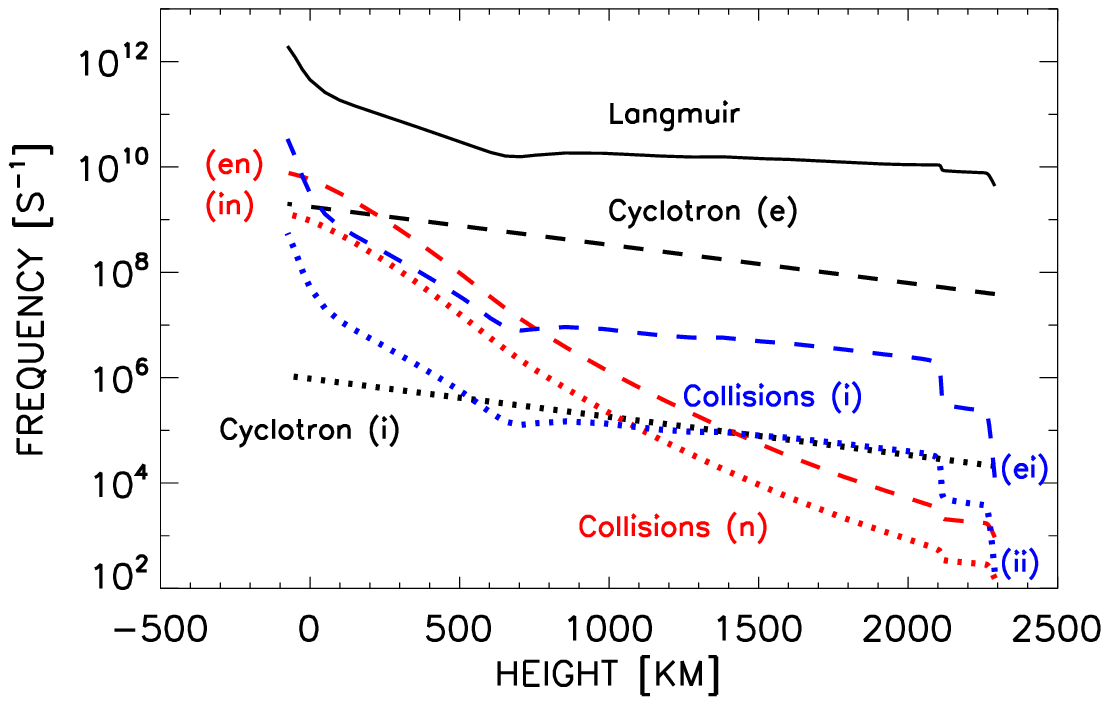}
\includegraphics[width=9cm]{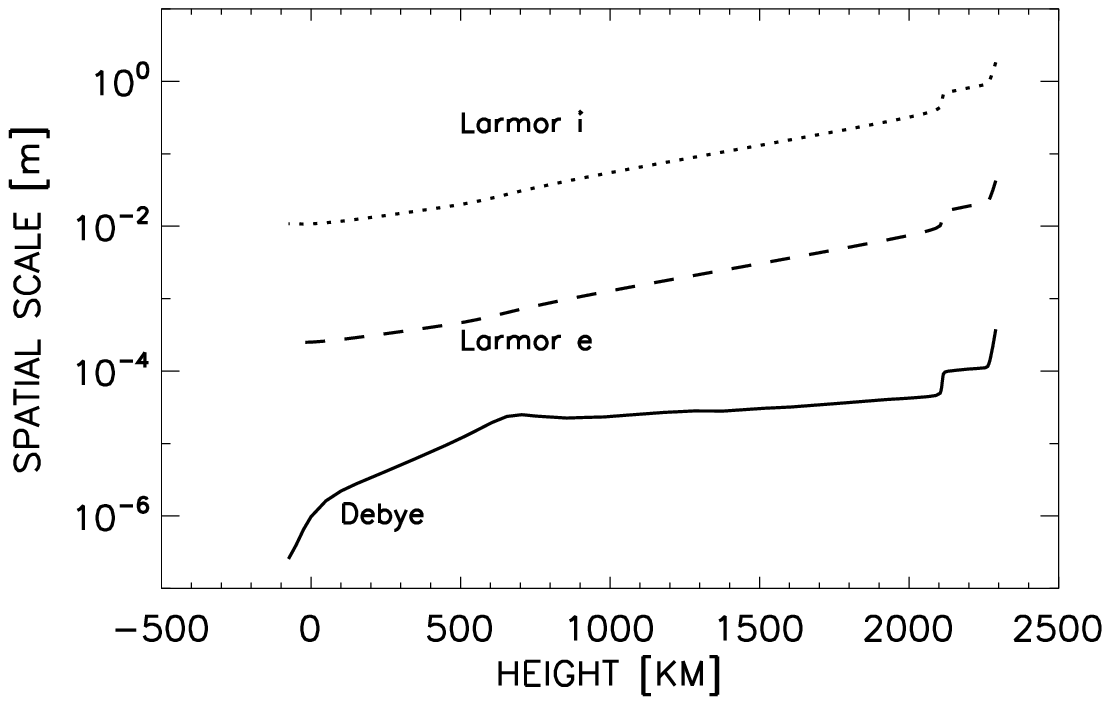}
\caption{{\footnotesize Plasma scales for the solar photosphere and chromosphere. {\it Upper panel:} collisions between neutral hydrogen with electrons (red dashed, Eq.~\ref{nu_en}) and with ions (red dotted, Eq.~\ref{nu_in}); collisions between hydrogen ions with electrons (blue dashed, Eq.~\ref{nu_ei}) and with themselves (blue dotted, Eq.~\ref{nu_ii}); cyclotron frequencies of gyration (black dotted and dashed, for ions and electrons, Eq.~\ref{eq:cyclotron}) for the field $B=100\exp{(-z/600)}$ G; and Langmuir frequency (black solid, Eq.~\ref{eq:langmuir}. {\it Lower panel:} Debye radius (black solid, Eq.~\ref{eq:debye}); Larmor radius of gyration for electrons and ions (black dotted and dashed, Eq.~\ref{eq:larmor}).}
}\label{fig:scales}
\end{figure}

\subsection{Assumptions of macroscopic equations}

The system of equations (Eqs.~\ref{eq:continuity-aie}, \ref{eq:momentum-aie}, \ref{eq:energy-aie}, and \ref{eq:rte_standard}) is so far rather general, but not closed, since the equations for the third momentum of the distribution functions are not present. The effect of thermal motions, of random phase cyclotron motions are averaged in these equations and are present in the form of kinetic pressure tensor (also called stress tensor), ${\bf\hat{p}}_{\alpha \mss{IE}}$, heat flow vector $\vec{q}_{\alpha \mss{IE}}$, and elastic and inelastic collisional terms $S_{\alpha \mss{IE}}$, $\vec{R}_{\alpha \mss{IE}}$, and $M_{\alpha \mss{IE}}$. To close the system, one has to provide the expressions for these quantities.

The pressure tensor can be approximated by a scalar pressure (only the elements of the diagonal are non-zero) in essentially two cases: (1) when the plasma is isotropic and collision-dominated, implying small mean free paths, and (2) for strongly magnetized plasmas when the cyclotron motion dominates over the collisions and the plasma becomes anisotropic in the direction perpendicular to the magnetic field, implying small gyration radius, much smaller that any gradient in any variable \citep[see, e.g.][]{Spitzer1956}. In the first case all three elements on the diagonal are the same. In the second case, there are two independent components of pressure, parallel and perpendicular to the magnetic field. The non-diagonal components of the pressure tensor are the origin of viscosity.

Radiative processes are considered to happen instantaneously, while other processes have characteristic temporal or spatial scales. Typical values used to determine the scale of the different physical mechanisms are the following:

\begin{itemize}

\item Frequency of Langmuir oscillations arising when the local charge neutrality of plasma is violated:
\begin{equation}
\omega_{pe}=(e^2n_e/m_e\epsilon_0)^{1/2} \label{eq:langmuir}
\end{equation}

\item Spatial scale of Langmuir  oscillations is the Debye radius
\begin{equation}
r_{pe}=(k_BT_e\epsilon_0/e^2n_e)^{1/2}
\label{eq:debye}
\end{equation}

\item Frequency of cyclotron rotation of particles of each species
\begin{equation}
\omega_{c\alpha \mss{I}}=|q_{\alpha \mss{I}}|B/m_{\alpha \mss{I}}, \,\,\, \mathsc{I} \neq 0 \label{eq:cyclotron}
\end{equation}

\item  Spatial scale  of cyclotron rotation is Larmor radius
\begin{equation}
r_{c\alpha  \mss{I}}=(2k_BT_{\alpha  \mss{I}}m_{\alpha  \mss{I}})^{1/2}/|q_{\alpha  \mss{I}}|B.
\label{eq:larmor}
\end{equation}

\end{itemize}

These frequencies and scales for typical parameters of the solar atmosphere are given in Fig.~\ref{fig:scales}, calculated after the VAL-C atmospheric model \citep{valc} and assuming a magnetic field varying with height as $B=100\exp{(-z/600)}$ G, approximating quiet solar regions. The largest spatial scale is the Larmor radius of ions reaching about 1 meter in the upper chromosphere. The lowest frequency is the ion cyclotron one reaching 10$^4$ Hz in the chromosphere (corresponding to a period of $2 \pi\times 10^{-4}$ sec).

The plasma frequencies and spatial scales defined above should be compared with the frequency of collisions between the charged particles and collisions between the neutral and charged particles. The frequency of collisions between neutral atoms of specie $\beta$ with ions of specie $\alpha$, and electrons are given in, e.g., \citet{Spitzer1956}, although some newer developments can be found in \citet{Vranjes2013}:
\begin{equation} \label{nu_in}
\nu_{i_\alpha n_\beta} = n_{\beta n}\sqrt{\frac{8 k_B T}{\pi m_{i_\alpha n_\beta}}}\Sigma_{in}
\end{equation}
\begin{equation} \label{nu_en}
\nu_{en_\beta} = n_{\beta n}\sqrt{\frac{8 k_B T}{\pi m_{i_\alpha n_\beta}}}\Sigma_{en}
\end{equation}
where $m_{i_\alpha n_\beta}=m_{i_\alpha} m_{n_\beta}/(m_{i_\alpha} + m_{n_\beta})$ and $\Sigma_{in}=5\times10^{-19}$ m$^2$, $\Sigma_{en}=10^{-19}$ m$^2$ are the ion-neutral and the electron-neutral cross sections, respectively.

The collisional frequencies of electrons with ions of specie $\alpha$, and of ions of different species are provided by \citet{Braginskii1965}, his Appendix A1 (see also \citet{Lifschitz, Rozhansky} and \citet{Bittencourt}, chapter 22):
\begin{equation} \label{nu_ei}
\nu_{ei_\alpha} = \frac{e^4 n_{\alpha i} \Lambda}{3\epsilon_0^2 m_e^2}\left(\frac{m_e}{2\pi k_B T}\right)^{3/2} = 3.7\times 10^{-6}\frac{n_{\alpha i} \Lambda}{T^{3/2}}
\end{equation}
\begin{equation} \label{nu_ii}
\nu_{i_\beta i_\alpha} = \frac{e^4 n_{\alpha i} \Lambda Z_{\alpha i}^4}{3\sqrt{2}\epsilon_0^2 m_{i_\beta}^2}\left(\frac{m_{i_\beta}}{2\pi k_B T}\right)^{3/2} = 6\times 10^{-8}\frac{n_{\alpha i}\Lambda Z_{\alpha i}^4}{T^{3/2}}
\end{equation}

\noindent where $Z_{i_\alpha}$ is the charge of the ion $\alpha$ ($Z_{i \alpha}=1, 2, ...$), and the mass of the ion is assumed to be approximately equal to the proton mass in evaluating the coefficient of the last equality.  The Coulomb logarithm $\Lambda$ for $T_e<50$ eV (i.e. solar temperatures approximately below 600.000 K) is
\begin{equation}
\Lambda=23.4 - 1.15\log_{10}{n_e}+3.45\log_{10}{T_e}
\label{Coulomb_log}
\end{equation}
\noindent with $n_e$ given in cm$^{-3}$ and $T$ in K. These frequencies for the solar atmosphere are plotted in red in Fig.~\ref{fig:scales}. The thermal velocities of ions are taken to be of the order of 10$^1$ \kms\ and those of electrons of the order of 10$^3$ \kms.

The quantities ${\bf\hat{p}}_{\alpha \mss{IE}}$, $\vec{q}_{\alpha \mss{IE}}$, $S_{\alpha \mss{IE}}$, $\vec{R}_{\alpha \mss{IE}}$, and $M_{\alpha \mss{IE}}$ can be expressed in terms of $\rho_{\alpha \mss{IE}}$, $p_{\alpha \mss{IE}}$, $u_{\alpha \mss{I}}$ and the magnetic field direction $\vec{b}=\vec{B}/B$ if the plasma temporal scale, $\tau$, and the spatial scales, $L_{\parallel}$ and $L_{\perp}$ (parallel and perpendicular to the magnetic field), satisfy the following conditions \citep{Lifschitz}:
\begin{eqnarray}
\label{eq:conditions}
\tau &  \gg & \omega_{c\alpha}^{-1} \\
\nonumber
\tau & \gg  & \nu_{ee}^{-1} \,, \nu_{ii}^{-1} \\ \nonumber
L_{\parallel} & \gg & \nu_{\alpha}^{-1}c_{\alpha} \\ \nonumber
L_{\parallel} & \gg & \nu_{\alpha}^{-1}u_{\alpha} \\ \nonumber
L_{\perp} & \gg & \omega_{c\alpha}^{-1}c_{\alpha} \\ \nonumber
L_{\perp} & \gg & \omega_{c\alpha}^{-1}u_{\alpha}
\end{eqnarray}
The temporal scales should be much larger than the cyclotron period and the time between collisions. The spatial scales parallel to the magnetic field should be much larger than those defined by the collision frequencies and the thermal and plasma velocities (i.e. thermal free path scale). In the case of scales perpendicular to the magnetic field, those should be larger than the dimensions defined by the cyclotron frequency and the velocities (i.e. radius of cyclotron gyration around the magnetic field lines). Fig.~\ref{fig:scales} shows that all these conditions are satisfied in the solar atmosphere, for the typical observed spatial and temporal scales.

It is important to note that the motion determined from macroscopic equations does not agree with the microscopic particle drifts, described by the equation of motion of individual particles. Only the drift due to electric field remains in the macroscopic equations. The drifts due to the gradient of the magnetic field is not present in the macroscopic description. The latter drift is responsible for, e.g., the motion of charged particles in the Earth's magnetosphere (magnetic mirrors), and is usually described in the microscopic approach. This apparent paradox in the description of plasma drifts and its origin is discussed in \citet{Spitzer1956} and has to do with the different way of averaging in the macroscopic and microscopic equations.

\section{Two-fluid description}

The solar plasma is composed by particles of different atomic elements in different ionization stages. One can consider that, once the collision coupling weakens, all these species behave in a different way, moving with different velocities. However, as a first approximation, it is reasonable to assume that the difference in behavior between neutrals and ions is larger than between the neutrals/ions of different kind themselves, since the latter feel the presence of the magnetic field and the former do not. This assumption allows to decrease the number of equations for different species $\alpha$ (Eq.~\ref{eq:continuity-aie}, \ref{eq:momentum-aie} and \ref{eq:energy-aie}) to just two, for an average neutral particle and an average charged particle (photons considered apart).  The two-fluid description has the limitation that it requires a stronger coupling between charged particles than between charged and neutral particles \citep[see,][]{Zaqarashvili2011b, Zaqarashvili2011} and thus it is only valid for z$>$1000 km according to Fig.\,\ref{fig:scales}. Also, in some circumstances, if the collisional coupling between the neutral species is weaker than between ionized and neutral ones, the neutral species can decouple as well one from another, see, e.g. \citet{Zaqarashvili2011b, Zaqarashvili2011}. The approach described in this section can be easily extended to describe this case. Below we derive the two-fluid equations for a plasma composed by an arbitrary number of neutral and ion species (not only hydrogen). The evolution of the charged species will be described by equations that combine jointly electrons and all ions. The evolution of neutral species will be described by the combined equations for all neutral components. Plasma quasi-neutrality is assumed and the temperature of electrons and ions is assumed the same (though the neutral component can have a different temperature). We assume that there are $N$ species present in the plasma, i.e. we have $N$ different ions and $N$ different neutrals ($2N+1$ species in total, accounting for electrons as well). We will only consider singly-ionized ions since the abundance of multiply-ionized ions is not large in the regions of interest of the solar atmosphere (i.e., regions where the presence of neutrals is important, photosphere and chromosphere). The conditions of the solar photosphere are such that the most abundant neutral atoms are hydrogen, and the donors of electrons are ionized iron atoms. The summation done in this section allows to obtain equation for an average neutral/ion atoms taking the latter feature into account.

In the next subsections, we use indices $n$ and $i$, instead of $\mathsc{I}=0$ or $\mathsc{I}=1$, to refer to neutral or ions, following the standard notation of previous works. Where appropriate, we also use the index $c$ to account for the combined effect of all charges, i.e, ions and electrons.

\subsection{Multi-species continuity equations}

\noindent Using Eq. \ref{eq:continuity-ai} and the definitions from Appendix~\ref{app:defs2} one trivially obtains:
\begin{equation} \label{eq:continuity-alfa-n}
\frac{\partial \rho_n}{\partial t} + \vec{\nabla} (\rho_n\vec{u}_n) = S_n
\end{equation}
\noindent In a similar way, for charges
\begin{equation} \label{eq:continuity-alfa-ie} \frac{\partial
\rho_c}{\partial t} + \vec{\nabla} (\rho_c\vec{u}_c) = S_i + S_e = S_c = -S_n
\end{equation}
\noindent where $\rho_c=\sum_\alpha \rho_{\alpha i} + \rho_e$, see Appendix~\ref{app:defs2}. The definitions of the inelastic collisional terms $S_n$, $S_i$ and $S_e$ are given in Appendix~\ref{app:SR}. The last equality comes from the fact that collisions can not change the total density.

\subsection{Photon momentum and energy equations for processes involving neutrals and charged species}

Before proceeding to the derivation of the momentum and energy equations for neutrals and charges, it is convenient to split the photon momentum and energy equations, Eqs.~\ref{eq:photon_momentum} and \ref{eq:photon_energy}, taking into account the nature of the interaction of particles and photons. In that equation, we can define the emissivity $j_\nu$ and absorption $k_\nu$ coefficients as the sum of separate contributions from neutrals (superindex $n$) and charges (superindex $c$):
\begin{equation}
j_\nu= j_\nu^n + j_\nu^c
\end{equation}
\begin{equation}
k_\nu= k_\nu^n + k_\nu^c
\end{equation}
Examples of this separation are: excitation(deexcitation) processes of neutral atoms will be included in $k_\nu^n(j_\nu^n)$, excitation/deexcitation of ionized species contribute to $k_\nu^c$/$j_\nu^c$; the ionization of a neutral species is taken into account in $k_\nu^n$ (a photon disappears from the radiation field due to the absorption by the neutral), the recombination of an ionized atom with an electron to give a neutral atom is included in $j_\nu^c$ (a photon appears as a consequence of the ion-electron interaction), etc.

With this, one can write
\begin{equation}
{\bf{\hat{P}}}_R = {\bf{\hat{P}}}_R^n + {\bf{\hat{P}}}_R^c
\end{equation}
where ${\bf{\hat{P}}}_R^n$ and ${\bf{\hat{P}}}_R^c$ are defined as
\begin{eqnarray} \label{eq:photon_momentum_neutrals} 
\vec{\nabla}{\bf{\hat{P}}}_R^n &=& \frac{1}{c} \int_0^{\infty} \oint (j_\nu^n - k_\nu^n I_\nu)\vec{n} d\Omega d\nu 
\end{eqnarray}
and
\begin{eqnarray} \label{eq:photon_momentum_charges}
\vec{\nabla}{\bf{\hat{P}}}_R^c &=& \frac{1}{c} \int_0^{\infty} \oint (j_\nu^c - k_\nu^c I_\nu)\vec{n} d\Omega d\nu 
\end{eqnarray}
Analogously, one can define
\begin{equation}
\vec{F}_R = \vec{F}_R^n + \vec{F}_R^c
\end{equation}
where 
\begin{eqnarray} \label{eq:photon_energy_neutrals}
\vec{\nabla}\vec{F}_R^n &=&  \int_0^{\infty} \oint (j_\nu^n - k_\nu^n I_\nu) d\Omega d\nu 
\end{eqnarray}
and
\begin{eqnarray} \label{eq:photon_energy_charges}
\vec{\nabla}\vec{F}_R^c &=&  \int_0^{\infty} \oint (j_\nu^c - k_\nu^c I_\nu) d\Omega d\nu 
\end{eqnarray}
and intensities $ I_\nu$ are obtained from complete radiative transfer equation, Eq. \ref{eq:rte_standard} with total coefficients $j_\nu$ and $k_\nu$.

\subsection{Multi-species momentum equation for neutrals}

First we start with neutrals, adding Eq. \ref{eq:momentum-ai} for $\alpha$ neutral species:
\begin{eqnarray}
\sum_{\alpha}\rho_{\alpha n}\frac{D\vec{u}_{\alpha n}}{Dt} &=&
\sum_{\alpha}\rho_{\alpha n}\vec{g}  - \sum_{\alpha}\vec{\nabla}{\bf\hat{p}}_{\alpha n} \nonumber \\
&+&\sum_{\alpha}\vec{R}_{\alpha n} - \sum_{\alpha}\vec{u}_{\alpha n}S_{\alpha n}
\end{eqnarray}

\noindent Using the continuity equation, we expand the derivative part as usual:
\begin{eqnarray}
\label{eq:lhs}
\sum_{\alpha}\rho_{\alpha n}\frac{D\vec{u}_{\alpha n}}{Dt}  + \sum_{\alpha}\vec{u}_{\alpha n}S_{\alpha n} &=&  \frac{\partial (\rho_n\vec{u_n})}{\partial t} + \vec{\nabla}(\rho_n\vec{u_n} \otimes \vec{u_n}) \nonumber \\
&+& \vec{\nabla}\sum_{\alpha}\rho_{\alpha n}\vec{w}_{\alpha n} \otimes \vec{w}_{\alpha n} \nonumber
\end{eqnarray}

\noindent The pressure tensor term gives:
\begin{equation}
\vec{\nabla}\sum_{\alpha}{\bf\hat{p}}_{\alpha n} = \vec{\nabla}{\bf\hat{p}}_n - \vec{\nabla}\sum_{\alpha}\rho_{\alpha n}\vec{w}_{\alpha n}\otimes\vec{w}_{\alpha n}
\end{equation}
where the drift velocity $\vec{w}_{\alpha n}$ is defined according to Eq.~\ref{eq:wie-alpha}. The terms containing  $\vec{w}_{\alpha n}$ cancel out after the summation of the above two expressions.

The collisional term takes into account the elastic collisions between all neutral species with ions and electrons (denoted by $\vec{R}_n$), as well as inelastic collisions through the momentum carried by photons (given by the integral below). The collisions between different pairs of neutrals go away after summation (see Appendix, Eq.~\ref{eq:R_n}):
\begin{equation}
\sum_{\alpha}\vec{R}_{\alpha n}=\vec{R}_n + \frac{1}{c} \int_0^{\infty} \oint (k_\nu^n I_\nu - j_\nu^n)\vec{n} d\Omega d\nu 
\end{equation}

In a photon absorption/emission process, the neutral atoms gain/lose the momentum given by the second term on the right-hand side of the equation.

Finally, taking into account Eq.~\ref{eq:photon_momentum_neutrals}, the momentum equation for the neutral component becomes:
\begin{equation} \label{eq:momentum-alfa-n-noS}
\frac{\partial (\rho_n\vec{u_n})}{\partial t} + \vec{\nabla}(\rho_n\vec{u_n} \otimes \vec{u_n})= \rho_n\vec{g} -\vec{\nabla}{\bf\hat{p}}_n-\vec{\nabla}{\bf\hat{P}}_R^n+\vec{R}_n
\end{equation}
\noindent or, alternatively, using the continuity equation this can be rewritten as:
\begin{equation} \label{eq:momentum-alfa-n}
\rho_n\frac{D \vec{u_n}}{\partial t} = \rho_n\vec{g} -\vec{\nabla}{\bf\hat{p}}_n -\vec{\nabla}{\bf\hat{P}}_R^n +\vec{R}_n +\vec{u}_n S_n
\end{equation}

\subsection{Multi-species momentum equation for charges}

The charged component momentum equations are (Eq. \ref{eq:momentum-ai} and \ref{eq:momentum-e}):
\begin{equation}
\rho_e\frac{D\vec{u}_e}{Dt} = -en_e(\vec{E} + \vec{u}_e\times\vec{B}) + \rho_e\vec{g} - \vec{\nabla} {\bf\hat{p}}_e + \vec{R}_e - \vec{u}_e S_e \nonumber
\end{equation}
\begin{eqnarray}
\sum_{\alpha}\rho_{\alpha i}\frac{D\vec{u}_{\alpha i}}{Dt}& =& \sum_{\alpha}q_{\alpha i}n_{\alpha i}(\vec{E} + \vec{u}_{\alpha i}\times\vec{B}) + \sum_{\alpha}\rho_{\alpha i}\vec{g} \nonumber \\
&-& \sum_{\alpha}\vec{\nabla}{\bf\hat{p}}_{\alpha i}  +  \sum_{\alpha}\vec{R}_{\alpha i} - \sum_{\alpha}\vec{u}_{\alpha i}S_{\alpha i} \nonumber
\end{eqnarray}
\noindent The sum of the derivative terms gives, same as before
\begin{eqnarray}
\sum_{\alpha}\rho_{\alpha i}\frac{D\vec{u}_{\alpha i}}{Dt} + \rho_e\frac{D\vec{u}_e}{Dt} + \sum_{\alpha}\vec{u}_{\alpha i}S_{\alpha i} + \vec{u}_e S_e=\nonumber \\
\frac{\partial (\rho_c\vec{u_c})}{\partial t} + \vec{\nabla}(\rho_c\vec{u}_c\otimes\vec{u}_c)+ \nonumber \\
\vec{\nabla}\sum_{\alpha}\rho_{\alpha i}(\vec{w}_{\alpha i} \otimes \vec{w}_{\alpha i}) + \vec{\nabla}\rho_e(\vec{w}_e\otimes \vec{w}_e) \nonumber
\end{eqnarray}
\noindent The Lorentz force term
\begin{equation}
\sum_{\alpha}q_{\alpha i}n_{\alpha i}(\vec{E} + \vec{u}_{\alpha i}\times\vec{B}) -  en_e(\vec{E} + \vec{u}_e\times\vec{B})=[\vec{J}\times\vec{B}]
\end{equation}
according to the definition of the current density, $J$ given by Eq.~\ref{eq:j-alpha}.
\noindent The pressure term
\begin{eqnarray}
\sum_{\alpha}\vec{\nabla}{\bf\hat{p}}_{\alpha i}+\vec{\nabla} {\bf\hat{p}}_e =  \vec{\nabla} {\bf\hat{p}}_{ie}
&-&\sum_{\alpha}\vec{\nabla}\rho_{\alpha i}(\vec{w}_{\alpha i}\otimes\vec{w}_{\alpha i})\nonumber \\
&-&\vec{\nabla}\rho_e(\vec{w}_e\otimes\vec{w}_e)
\end{eqnarray}
according to Eq.~\ref{eq:pipe-alpha}.

The collision term writes as
\begin{eqnarray}
\sum_{\alpha}\vec{R}_{\alpha i} + \vec{R}_e & = & \vec{R}_{ie} + \int_0^{\infty} \oint (k_\nu^c I_\nu - j_\nu^c) d\Omega d\nu \nonumber \\
& = & -\vec{R}_n + \int_0^{\infty} \oint (k_\nu^c I_\nu - j_\nu^c) d\Omega d\nu
\end{eqnarray}
where the elastic collisions between ions and electrons with neutrals are taken into account (term $\vec{R}_e$) and the collisions between different pairs of ions or ions-electrons go away (see Appendix, Eq.~\ref{eq:R_e} and \ref{eq:R_i}). The integral term gives the momentum exchange due to ineslatic collisions. The last equality is derived from the conservation of the total momentum by elastic collisions.

With all this, the ion-electron momentum equation becomes:
\begin{equation} \label{eq:momentum-alfa-ie-noS}
\frac{\partial (\rho_c\vec{u_c})}{\partial t} + \vec{\nabla}(\rho_c\vec{u}_c\otimes\vec{u}_c)=[\vec{J}\times\vec{B}] + \rho_c\vec{g}-\vec{\nabla} {\bf\hat{p}}_{ie} -\vec{\nabla} {\bf\hat{p}}_{R}^c-\vec{R}_n
\end{equation}
Using the continuity equation, Eq.~\ref{eq:continuity-alfa-ie}, it can be rewritten as
\begin{equation} \label{eq:momentum-alfa-ie}
\rho_c\frac{D \vec{u}_c}{\partial t} =[\vec{J}\times\vec{B}] +  \rho_c\vec{g}-\vec{\nabla} {\bf\hat{p}}_{ie}-\vec{\nabla} {\bf\hat{p}}_{R}^c-\vec{R}_n+\vec{u}_cS_n
\end{equation}

\subsection{Multi-species energy equation for neutrals}

\noindent Adding up $\alpha$ energy equations for the neutral components (Eq.~\ref{eq:energy-p} and \ref{eq:energy-ion}, summed over $\mathsc{E}$ ionization states), we get:
\begin{eqnarray} \label{eq:energy-n-start}
&&\sum_{\alpha}\left(\frac{3}{2}\frac{D p_{\alpha n} }{Dt} + \frac{3}{2}p_{\alpha n}\vec{\nabla}\vec{u}_{\alpha n}  + ({\bf\hat{p}}_{\alpha n}\vec{\nabla})\vec{u}_{\alpha n} + \vec{\nabla}\vec{q}_{\alpha n} \right) =\nonumber \\
&   &\sum_{\alpha}(M_{\alpha n} - \vec{u}_{\alpha n}\vec{R}_{\alpha n} + \frac{1}{2}u_{\alpha n}^2S_{\alpha n})
\end{eqnarray}
and
\begin{equation} 
\sum_{\alpha}\frac{D \chi_{\alpha n}}{D t}+ \chi_{\alpha n} \vec{\nabla}\vec{u}_{\alpha n} =\sum_{\alpha}\Phi_{\alpha n}
\end{equation}
where $\chi_{\alpha n}$ and $\Phi_{\alpha n}$ are defined by Eqs. \ref{eq:chi-ai} and \ref{eq:Phi-ai}. 

Using the momentum and continuity equation, as well as definitions from Appendix~\ref{app:defs2}, after some lengthy but straightforward calculations, one obtains the total energy equation for neutrals:
\begin{eqnarray}
\frac{\partial }{\partial t}\left(e_n+\frac{1}{2}\rho_n u_n^2\right)&+& \vec{\nabla}\left(\vec{u}_n (e_n+\frac{1}{2}\rho_n u_n^2) \right) + \vec{\nabla}({\bf\hat{p}}_n\vec{u}_n)  \nonumber \\
+ \vec{\nabla}\vec{q}_n^{\prime}=\rho_n\vec{u}_n\vec{g} &+& \sum_{\alpha}M_{\alpha n}+\sum_{\alpha}\Phi_{\alpha n}\end{eqnarray}
where
\begin{equation}
e_n = 3p_n/2 + \sum_\alpha \chi_{\alpha n} = 3p_n/2 + \chi_n
\end{equation}
where $\vec{q}_n^{\prime}$ is defined according to Eq. \ref{eq:qie-alpha-prime}.

As before we can use momentum and continuity equations for neutrals to remove the kinetic energy part from this equation, getting:
\begin{eqnarray}
\label{eq:energy-alfa-n}
\frac{D e_n}{D t} + e_n\vec{\nabla}\vec{u}_n + ({\bf\hat{p}_n}\vec{\nabla})\vec{u}_n + \vec{\nabla}\vec{q}_n^{\prime} + \vec{\nabla}\vec{F}_R^n = \nonumber \\
M_n +\frac{1}{2}u_n^2 S_n-\vec{u}_n\vec{R}_n
\end{eqnarray}
where we have used the definition
\begin{eqnarray}
\sum_{\alpha}M_{\alpha n} &+& \sum_{\alpha}\Phi_{\alpha n}
=  \\
M_n &+& \int_0^{\infty} \oint (k_\nu^n I_\nu - j_\nu^n)\vec{n} d\Omega d\nu = M_n - \vec{\nabla}\vec{F}_R^n \nonumber
\end{eqnarray}
The term $M_n$ accounts for the energy exchange of neutrals through the elastic collisions with charges and the integral represents the energy gain/losses of neutral species after absorbing/emitting photons.

\subsection{Multi-species energy equation for charges}

Using the $\alpha$ energy equations for ions (Eq.~\ref{eq:energy-p} and \ref{eq:energy-ion}) and Eq.~\ref{eq:energy-e} for electrons, we get:
\begin{eqnarray} \label{eq:energy-i-start}
\frac{3}{2}\sum_{\alpha}\frac{D p_{\alpha i} }{Dt} &+&  \frac{3}{2}\sum_{\alpha}p_{\alpha i}\vec{\nabla}\vec{u}_{\alpha i} +
\sum_{\alpha}({\bf\hat{p}}_{\alpha i}\vec{\nabla})\vec{u}_{\alpha i} + \nonumber \\
\sum_{\alpha}\vec{\nabla}\vec{q}_{\alpha i} &=& \sum_{\alpha}(M_{\alpha i} - \vec{u}_{\alpha i}\vec{R}_{\alpha i} +  \frac{1}{2}u_{\alpha i}^2S_{\alpha i})
\end{eqnarray}
\begin{eqnarray} \label{eq:energy-e-start}
\frac{D}{Dt}\frac{3p_e}{2} +  \frac{3}{2}p_e\vec{\nabla}\vec{u}_e + ({\bf\hat{p}_e}  \vec{\nabla})  \vec{u}_e + \vec{\nabla}  \vec{q}_e  =\nonumber \\
M_e - \vec{u}_e  \vec{R}_e + \frac{1}{2}u_e^2S_e
\end{eqnarray}
and
\begin{equation} 
\sum_{\alpha}\frac{D \chi_{\alpha i}}{D t}+ \chi_{\alpha i} \vec{\nabla}\vec{u}_{\alpha i} =\sum_{\alpha}\Phi_{\alpha i}
\end{equation}
with $\chi_{\alpha i}$ and $\Phi_{\alpha i}$ are defined by Eqs. \ref{eq:chi-ai} and \ref{eq:Phi-ai}. 

\noindent In a similar way as for neutrals, after some lengthy manipulations the energy equation for charges is obtained:
\begin{eqnarray}
\frac{\partial }{\partial t}\left(e_{ei}+\frac{1}{2}\rho_c u_c^2\right) &+& \vec{\nabla}\left(\vec{u}_c(e_{ei}+\frac{1}{2}\rho_c u_c^2 ) \right) + \vec{\nabla}({\bf\hat{p}}_{ie}\vec{u}_c) + \nonumber \\
\vec{\nabla}\vec{q}_{ie}^{\prime} =  \rho_c\vec{u}_c\vec{g} &+& \vec{J}\vec{E} +\sum_{\alpha}M_{\alpha i} + M_e+\sum_{\alpha}\Phi_{\alpha i}
\end{eqnarray}
where
\begin{equation}
e_{ei} = 3p_{ei}/2 + \sum_\alpha \chi_{\alpha i} = 3p_{ei}/2 + \chi_i
\end{equation}
and $\vec{q}_{ie}^{\prime}$ is defined according to Eq. \ref{eq:qie-alpha-prime}.

We use the continuity and momentum equations for charges (Eq.~\ref{eq:continuity-alfa-ie} and \ref{eq:momentum-alfa-ie}) to remove the  kinetic energy part from this equation:
\begin{eqnarray}
\label{eq:energy-alfa-ie}
\frac{D e_{ie}}{D t} + e_{ie}\vec{\nabla}\vec{u}_c + ({\bf\hat{p}}_{ie}\vec{\nabla})\vec{u}_c + \vec{\nabla}\vec{q}_{ie}^{\prime} &+& \vec{\nabla}\vec{F}_R^c =   \nonumber \\
\vec{J}(\vec{E}+[\vec{u}_c\times\vec{B}]) - M_n -\frac{1}{2}u_c^2 S_n&+&\vec{u}_c\vec{R}_n 
\end{eqnarray}
where we have used the relation
\begin{equation}
\sum_{\alpha}M_{\alpha i} +\sum_{\alpha}\Phi_{\alpha i} + M_e = -M_n -\vec{\nabla}\vec{F}_R^c,
\end{equation}
due to the energy conservation in elastic collisions and the energy exchange with photons.
 
The derived system of equations (Eqs.~\ref{eq:continuity-alfa-n}, \ref{eq:continuity-alfa-ie}, \ref{eq:momentum-alfa-n}, \ref{eq:momentum-alfa-ie}, \ref{eq:energy-alfa-n} and \ref{eq:energy-alfa-ie}) looks formally the same as in the case of purely hydrogen plasma. However the variables are defined in the different way and represent average quantities over all N components of the plasma.  It is also important to note that only the collisions with or by neutrals need to be considered.


\begin{figure}
\center
\includegraphics[width=8cm]{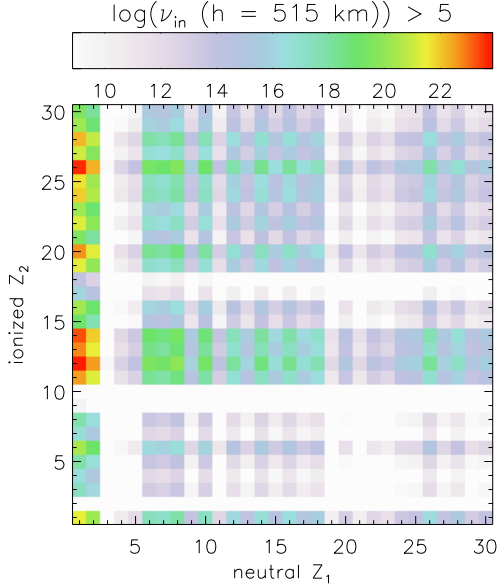}
\caption{{\footnotesize  Map of the collisional frequencies computed according to Eqs.~\ref{nu_en} and \ref{nu_in} at $h=515$~km of the VAL-C model atmosphere. The x-axis is the atomic number of the first collider, the y-axis the atomic number of the second collider.}
}\label{fig:collisions}
\end{figure}

\subsection{Collisional $\vec{R}$ terms}
\label{sec:collterms}

In order to make practical use of the collisional terms $\vec{R}$ in the momentum equations one has to express them via the average velocity of charges and neutrals instead of individual velocities as in Eqs. \ref{eq:R_i}, \ref{eq:R_n} and \ref{eq:R_e}. To that aim, we use the particular form of the collisional frequencies, Eqs.~\ref{nu_in}, \ref{nu_en} and \ref{nu_ei}. These frequencies essentially depend on the number of colliding particles, their masses, and a given function of temperature. Assuming we have collisions between electrons, ions of type $\alpha$ and neutrals of type $\beta$, one can write:
\begin{eqnarray} \label{eq:nus-approximated}
\nu_{ei_\alpha} &=& n_{\alpha i}m_e^{-1/2}f(T) \nonumber \\
\nu_{i_\alpha n_\beta} &=& n_{\beta n } m_{i_\alpha n_\beta}^{-1/2}\tilde{f}_1(T) = n_{\beta n } m_p^{-1/2}\tilde{f}_1(T)A_{\alpha\beta}^{-1/2} \nonumber \\
\nu_{e n_\beta} &=& n_{\beta n } m_e^{-1/2}\tilde{f}_2(T) 
\end{eqnarray}
where $m_p$ is the proton mass, $A_\alpha$ is the atomic mass of species $\alpha$, and $A_{\alpha\beta}=A_\alpha A_\beta/(A_\alpha + A_\beta)$ is the reduced mass.

Collisional frequencies satisfy the following property:
\begin{eqnarray}
\rho_{\alpha i}\nu_{i_\alpha n_\beta} &=&\rho_{\beta n}\nu_{n_\beta i_\alpha} \\
\rho_e\nu_{en_\beta}&=& \rho_{\beta n}\nu_{n_\beta e} \\
\rho_{\alpha i}\nu_{i_\alpha e} &=& \rho_e\nu_{ei_\alpha}
\end{eqnarray}

Using these properties we can rewrite the ion collisional term $\vec{R}_i$, Eq.~\ref{eq:R_i}, as:
\begin{eqnarray}
&&\vec{R}_i=-\rho_e m_e^{-1/2}f(T)\sum_{\alpha=1}^N n_{\alpha i}(\vec{u}_{\alpha i}- \vec{u}_e) - \nonumber \\
&-&m_p^{-1/2}\tilde{f}_1(T)\sum_{\alpha=1}^N\rho_{\alpha i}\sum_{\beta=1}^Nn_{\beta n}A_{\alpha\beta}^{-1/2} (\vec{u}_{\alpha_i} - \vec{u}_{\beta n}) \approx \nonumber \\
&-&\frac{\rho_e m_e^{-1/2}f(T)}{e}\vec{J} - m_p^{-1/2}\tilde{f}_1(T)n_n \rho_i (\vec{u}_i - \vec{u}_n) \approx \nonumber \\
&-& \frac{m_e}{e}\vec{J} \sum_{\alpha=1}^N \nu_{ei_\alpha} - \rho_i(\vec{u}_i - \vec{u}_n) \sum_{\alpha=1}^N\sum_{\beta=1}^N\nu_{i_\alpha n_\beta}
\end{eqnarray}

\noindent In a similar way we get for the neutral term, $\vec{R}_n$, Eq.~\ref{eq:R_n}:
\begin{eqnarray}
\vec{R}_n \approx &-&\rho_e(\vec{u}_n - \vec{u}_e) \sum_{\beta=1}^N \nu_{en_\beta}\nonumber \\
&-&\rho_i(\vec{u}_n - \vec{u}_i) \sum_{\alpha=1}^N\sum_{\beta=1}^N\nu_{i_\alpha n_\beta}
\end{eqnarray}

\noindent and for the electron term, $\vec{R}_e$, Eq.~\ref{eq:R_e}:
\begin{eqnarray}
\vec{R}_e \approx \sum_{\alpha=1}^N \nu_{ei_\alpha}\frac{m_e}{e}\vec{J} + \sum_{\beta=1}^N \nu_{en_\beta}\rho_e(\vec{u}_n - \vec{u}_e)
\end{eqnarray}
where we have neglected the term $\sum_\beta n_{\beta n}\vec{w}_{\beta n}$, since we do not expect large differences between the individual velocities of neutrals of type $\beta$ with respect to their center of mass velocity. Therefore, we assume the magnitude of this term to be of second order. Additionally, we have assumed that the factor $A_{\alpha\beta}^{-1/2}$ does not differ significantly from its value for hydrogen atoms. This number is only large for collisions between pairs of heavy particles, with large atomic number, as e.g., Fe. Since the abundance of these atoms in the solar atmosphere is significantly smaller that of the lightest atoms (H, He), we can safely neglect their contribution in the summation.

To prove the latter idea we have calculated the frequency of collisions according to Eqs.~\ref{nu_in} and \ref{nu_en}, multiplied by the number density of both colliders, i.e.,
\begin{equation}
\tilde{\nu}_{i_\alpha n_\beta}=n_{\alpha i}n_{\beta n}\sqrt{\frac{8 k_B T}{\pi m_p A_{\alpha\beta}}}\Sigma_{in}
\end{equation}
We have computed $\nu_{i_\alpha n_\beta}$ for the first 30 elements of the periodic system and their first ions. For that, we have used their solar relative abundances and have computed the Saha distribution of population densities for the first-ionized states of those 30 elements. Then we have solved the equations of the instantaneous chemical equilibrium for the 30 elements and 14 molecules that are common in the solar atmosphere (CH, NH, OH, CN, CO, NO, TiO, H$_2$, H$_2^+$ , C$_2$, N$_2$, LiH, SiH and HF). Finally, we have computed the collisional frequencies. The computation has been performed for the electron density, temperature and total hydrogen density of the VALC model \citep{valc}.

The result of these calculation is given in Fig.~\ref{fig:collisions}. We have chosen as representative an arbitrary height in the upper photosphere, 515 km. The relative weight between the collisional frequencies does not significantly depend on height. It appears that the most frequent collisions are between the most abundant elements, H and He, with ions of other elements. The rest of colliders are much less abundant and therefore, the collisional transfer of momentum between them can be safely neglected. Therefore, we can safely assume $A_{\alpha\beta}^{-1/2}$ only weakly varying, and perform the summation above.

\parskip 0pt

\subsection{Ohm's law for two-fluid description}

\parskip 0pt

Ohm's law is frequently derived from the electron momentum equation, by neglecting the electron inertia (terms ${\partial (\rho_e\vec{u_e})}/{\partial t}$, $\vec{\nabla}(\rho_e\vec{u_e} \otimes \vec{u_e})$) and gravity acting on electrons ($\rho_e\vec{g}$), see, e.g.  \citet{Zaqarashvili2011b}. However, we now deal with the $2N+1$ component plasma ($N$ ionized, $N$ neutral and 1 electron components) and we find more convenient to use the same strategy as in \cite{Bittencourt}. We derive an evolution equation for the electric current by multiplying the momentum equations (Eq.~\ref{eq:momentum-ai} and \ref{eq:momentum-e}) of each species by $q_{\alpha i}/m_{\alpha \mss{I}}$ and add them up. Since neutral species have zero charge, summatories only extend to ions ($\alpha=1,.., N$) and electrons ($\alpha=N+1$). This leads to
\begin{eqnarray} \label{eq:ohm-alfa-start}
&&\sum_{\alpha=1}^{N+1} \left( n_{\alpha i} q_{\alpha i} \frac{\partial \vec{u}_{\alpha i}}{\partial t} + n_{\alpha i} q_\alpha(\vec{u}_{\alpha i} \vec{\nabla})\vec{u}_{\alpha i} + \frac{q_{\alpha i}}{m_{\alpha i}} \vec{u}_{\alpha i} {S_{\alpha i}} \right) = \nonumber \\
&&\sum_{\alpha=1}^{N+1}\left(\frac{n_{\alpha i} q_ {\alpha i}^2}{m_{\alpha i}}(\vec{E} + \vec{u}_{\alpha i} \times \vec{B}) + n_{\alpha i} q_{\alpha i} \vec{g}-\vec{\nabla}\frac{q_{\alpha i}}{m_{\alpha i}} {\bf\hat{p}}_{\alpha i}  \right)+ \nonumber  \\
&&\sum_{\alpha=1}^{N+1}\frac{q_{\alpha i}}{m_{\alpha i}} \vec{R}_{\alpha i} 
\end{eqnarray}
Using the continuity equation for species $\alpha$, one gets:
\begin{eqnarray} \label{eq:intermediate3}
&&\sum_{\alpha=1}^{N+1} \left(n_{\alpha i} q_{\alpha i} \frac{\partial \vec{u}_{\alpha i}}{\partial t} + n_{\alpha i} q_{\alpha i} (\vec{u}_{\alpha i}  \vec{\nabla})\vec{u}_{\alpha i}  +  \frac{q_{\alpha i}}{m_{\alpha i}} \vec{u}_{\alpha i} S_{\alpha i} \right) = \nonumber \\
&&\frac{\partial \vec{J}}{\partial t} +  \vec{\nabla}\sum_{\alpha=1}^{N+1} ( n_{\alpha i} q_{\alpha i} \vec{u}_{\alpha i} \otimes \vec{u}_{\alpha i}) 
\end{eqnarray}
\noindent Here we have used the current density definition $\vec{J}=\sum_{\alpha}n_{\alpha i} q_{\alpha i} \vec{u}_{\alpha i}$ according to Eq.~\ref{eq:j-alpha}.

Expanding the convective term in Eq.~\ref{eq:intermediate3}, and substituting the velocity $\vec{u}_{\alpha i}$ using the definition of the relative velocities of ions and electrons (Eq.~\ref{eq:wie-alpha}) with respect to the center of mass velocity of charges $\vec{u}_c$ ($\vec{u}_{\alpha i}=\vec{w}_{\alpha i} + \vec{u}_c$ and $\vec{u}_e=\vec{w}_e + \vec{u}_c$) we have:
\begin{eqnarray}
\vec{\nabla}\sum_{\alpha=1}^{N+1}(n_{\alpha i} q_{\alpha i} \vec{u}_{\alpha i} \otimes \vec{u}_{\alpha i}) = \vec{\nabla}(\vec{J} \otimes \vec{u}_c) + \vec{\nabla}(\vec{u}_c\otimes \vec{J^\prime}) + \nonumber \\
\vec{\nabla}\sum_{\alpha=1}^N(n_{\alpha i}q_{\alpha i}\vec{w}_{\alpha i}\otimes \vec{w}_{\alpha i})  - e\vec{\nabla}(n_e\vec{w}_e\otimes\vec{w}_e) \nonumber
\end{eqnarray}

\noindent Here we have explicitly replaced species $\alpha=N+1$ by electrons.

The current density $\vec{J}^\prime$ is defined as
\begin{eqnarray}
\vec{J}^\prime = \sum_{\alpha=1}^N n_{\alpha i}q_{\alpha i}\vec{w}_{\alpha i} -en_e\vec{w}_e =\vec{J}
\end{eqnarray}
the later equality holds because the charge neutrality is assumed.

All four terms in the above expansion are of second-order, given that they involve double products of velocities.  Generally, the Ohm's law is used only up to first order, and they are neglected. We will do the same approximation later. We keep them for the moment, in case they need to be retained (partially or fully) for some reason. The gravity term in Eq.~\ref{eq:ohm-alfa-start} also cancels out because of the charge neutrality. This gives the Ohm's law the following form:
\begin{eqnarray} \label{eq:ohm-bit1}
\frac{\partial\vec{J}}{\partial t} &+& \vec{\nabla}(\vec{J}\otimes \vec{u}_c + \vec{u}_c \otimes \vec{J}) + \vec{\nabla} \sum_{\alpha=1}^N(n_{\alpha i}q_{\alpha i}\vec{w}_{\alpha i}\otimes\vec{w}_{\alpha i}) \nonumber \\
&-& e\vec{\nabla}(n_e\vec{w}_e\otimes\vec{w}_e) = \sum_{\alpha=1}^{N+1}\left( \frac{n_{\alpha i} {q_{\alpha i}}^2}{m_{\alpha i}}(\vec{E} + \vec{u}_{\alpha i} \times \vec{B}) \right) \nonumber \\
 &-& \sum_{\alpha=1}^{N+1}\left(\vec{\nabla}\frac{q_{\alpha i}}{m_{\alpha i}}{\bf\hat{p}}_{\alpha i}  + \frac{q_{\alpha i}}{m_{\alpha i}} \vec{R}_{\alpha i}\right)
\end{eqnarray}

\noindent Up to now it is general and no assumptions have been made except for the charge neutrality and the neglecting of multiply ionized atoms. To make a practical use of this Ohm's law we will particularize it to our case. Setting $q_\alpha=e$ for ions and putting explicitly the electron contribution, the Lorentz term sums to:
\begin{eqnarray}
&&\sum_{\alpha=1}^{N+1}\frac{n_{\alpha i}q_{\alpha i}^2}{m_{\alpha i}}(\vec{E} +
\vec{u}_\alpha \times \vec{B})= \nonumber \\
&&\frac{e^2n_e}{m_e}\left( \sum_{\alpha=1}^{N}\frac{n_{\alpha i}}{n_e}\frac{m_e}{m_{\alpha i}}
+ 1 \right)[\vec{E} + \vec{u}_c\times{B}]-\frac{e}{m_e}[\vec{J}\times\vec{B}] + \nonumber \\
&&\frac{e^2}{m_e}\left( \sum_{\alpha=1}^N n_{\alpha i}\vec{w}_{\alpha i}(1 + \frac{m_e}{m_{\alpha i}}) \right)\times\vec{B} 
\end{eqnarray}

We proceed by neglecting all the terms proportional to $m_e/m_{\alpha i}$ in the Lorentz force term. The drift velocity term is expanded as:
\begin{equation}
\sum_{\alpha=1}^Nn_{\alpha i}\vec{w}_{\alpha i}=\sum_{\alpha=1}^Nn_{\alpha i}\vec{u}_{\alpha i} - n_i\vec{u}_c
\end{equation}
The result of the summation can be reasonably assumed small in the case of multiple ions, since we do not expect strong deviations between the individual velocities of ions $\vec{u}_{\alpha i}$ and the center of mass velocity of charges $\vec{u}_c$. After neglecting this term, we get
\begin{eqnarray}
\sum_{\alpha=1}^{N+1}\frac{n_{\alpha i} q_{\alpha  i}^2}{m_{\alpha i}}(\vec{E} + \vec{u}_{\alpha i} \times \vec{B}) &\approx &
\frac{e^2n_e}{m_e}[\vec{E} + \vec{u}_c\times{B}] \nonumber  \\
&-&\frac{e}{m_e}[\vec{J}\times\vec{B}]
\end{eqnarray}

\noindent The sum of the pressure terms gives:
\begin{eqnarray} \label{eq:psum-ohm}
\vec{\nabla} \left[ \sum_{\alpha=1}^{N+1}\frac{q_{\alpha i}}{m_{\alpha i}} {\bf\hat{p}}_{\alpha i} \right] &=& \frac{e}{m_e}\vec{\nabla}\left( \sum_{\alpha=1}^N{\bf\hat{p}}_{\alpha i}\frac{m_e}{m_{\alpha i}}  - {\bf\hat{p}}_e\right) \approx  \nonumber  \\
&-&\frac{e\vec{\nabla}{\bf\hat{p}}_e}{m_e}
\end{eqnarray}

Next we deal with the friction forces terms. Note that because of the multiplication by $q_\alpha$, the friction force of neutrals has no effect. In order to operate these terms we have to take into account the form of expression for collisional frequencies, Eqs.~\ref{nu_en}, \ref{nu_in}, Eq.~\ref{nu_ei}, and Eq.~\ref{eq:nus-approximated}.
\begin{eqnarray} \label{eq:rsum-ohm}
\sum_{\alpha=1}^N\frac{e}{m_{\alpha \mss{I}}}\vec{R}_i &-& \frac{e}{m_e}\vec{R}_i \approx - \vec{J}\left(\sum_{\alpha=1}^N \nu_{ei_\alpha} + \sum_{\beta=1}^N \nu_{en_\beta} \right) \nonumber \\
+ en_e (\vec{u}_c &- &\vec{u}_n)\left( \sum_{\beta=1}^N\nu_{en_\beta} - \sum_{\alpha=1}^N\sum_{\beta=1}^N\nu_{i_\alpha n_\beta} \right)
\end{eqnarray}
where we have neglected terms such as $\sum_\alpha n_{\alpha i}\vec{w}_{\alpha i}$ and assumed weak variations of the factor $A_{\alpha\beta}^{-1/2}$, similar to Section~\ref{sec:collterms}.

After assuming stationary currents and neglecting second-order terms (all terms on the left hand side of Eq.~\ref{eq:ohm-bit1}), the two-fluid Ohm's law for multi-component plasma becomes:
\begin{eqnarray}
\vec{E}^*&=&[\vec{E} + \vec{u}_c\times{B}] = \frac{1}{en_e}[\vec{J}\times \vec{B}] -\frac{\vec{\nabla}{\bf\hat{p}}_e}{en_e} \nonumber \\
&+& \frac{\rho_e}{(en_e)^2}\left(\sum_\alpha \nu_{ei_\alpha} + \sum_\beta \nu_{en_\beta} \right)\vec{J}  \\
&-& \frac{\rho_e}{en_e}(\vec{u}_c - \vec{u}_n)\left( \sum_\beta\nu_{en_\beta} - \sum_\alpha\sum_\beta\nu_{i_\alpha n_\beta} \right) \nonumber
\end{eqnarray}

This equation closes the system. From left to right the terms in the Ohm's law are: Hall term; battery term, Ohmic term, and ambipolar term. The Ohm equation has similar form as the one for hydrogen plasma derived in e.g., \citet{Zaqarashvili2011} except that we use $\vec{u}_c$ instead of $\vec{u}_i$.

\section{Single-fluid description}
\label{app:single_fluid}

When the collisional coupling of the plasma is strong enough, it is convenient to use a single-fluid quasi-MHD approach  i.e, including resistive terms that are not taken into account by the ideal MHD approximation. The derivation of the single-fluid equations for a multi-component plasma ($N$ neutrals, $N$ ions and one electron component) goes essentially through the same steps as for two-fluids above, except for slightly different definitions of the macroscopic variables. These definitions are given in Appendix~\ref{app:defs1}. In this section we provide the final single-fluid equations of conservation of mass, momentum and energy, and derive the generalized Ohm's law for the case of single-fluid description. Below we use single index $\alpha$ to refer to each of $2N+1$ components, without explicit indication of neutrals and ions.

\subsection{Mass, momentum and energy conservation}

The mass, momentum and energy equations for all species, including photons, are added up leading to the single-fluid equations in the following form:
\begin{eqnarray} \label{eq:continuity-single}
\frac{\partial \rho}{\partial t} + \vec{\nabla}\left(\rho\vec{u}\right) =  \sum_{\alpha=1}^{2N+1} S_\alpha = 0
\end{eqnarray}
\begin{equation} \label{eq:momentum-single}
\rho\frac{D\vec{u}}{D t}  = \vec{J}\times\vec{B} + \rho\vec{g}  - \vec{\nabla}{\bf\hat{p}} - \vec{\nabla}{\bf\hat{P}}_R
\end{equation}
\begin{eqnarray} \label{eq:energy-single}
\frac{\partial }{\partial t}\left(e + \frac{1}{2}\rho u^2 \right) & + & \vec{\nabla}  \left( \vec{u}\, ( e + \frac{1}{2}\rho u^2) +{\bf\hat{p}}\vec{u} \right) + \nonumber \\ 
& +& \vec{\nabla}  \vec{q}^{\prime} + \vec{\nabla} \vec{F}_R  = \vec{J} \vec{E}   + \rho\vec{u} \vec{g}
\end{eqnarray}
where
\begin{equation}
e=\frac{3}{2}p + \sum_{\alpha=1}^{2N}\chi_\alpha
\end{equation}
and $\vec{q}^{\prime}$ defined by equation Eq. \ref{eq:q-prime} and $\chi_\alpha$ by Eq. \ref{eq:chi-ai} with the index $\alpha$ running over $N$ neutrons plus $N$ ions. 

In these equations we have not neglected the electron inertia. The individual inelastic collisional $S_\alpha$-terms have disappeared since there can not exist any mass density variation due to collisions. The momentum transfer between particles also vanishes (the elastic collision terms between particles sum to zero according to Eqs.~\ref{eq:R_e}, \ref{eq:R_i} and \ref{eq:R_n}) and only the momentum exchange with photons remains (through the term $-\vec{\nabla}{\bf\hat{P}}_R$). The same happens with the energy transfer between particles and only the energy exchange with photons is left with the term $\vec{\nabla} \vec{F}_R$.

We can remove the kinetic energy terms from the energy equation using the momentum (Eq.~\ref{eq:momentum-single}) and continuity (Eq.~\ref{eq:continuity-single}) equations, obtaining:
\begin{equation}
\label{eq:energy-single-p}
\frac{D e}{D t} + e \vec{\nabla}  \vec{u} + ({\bf\hat{p}}  \vec{\nabla})  \vec{u} + \vec{\nabla}  \vec{q}^{\prime} + \vec{\nabla} \vec{F}_R  = \vec{J}  [\vec{E} + \vec{u} \times \vec{B}] 
\end{equation}

\begin{figure*}[t]
\center
\includegraphics[width=7cm]{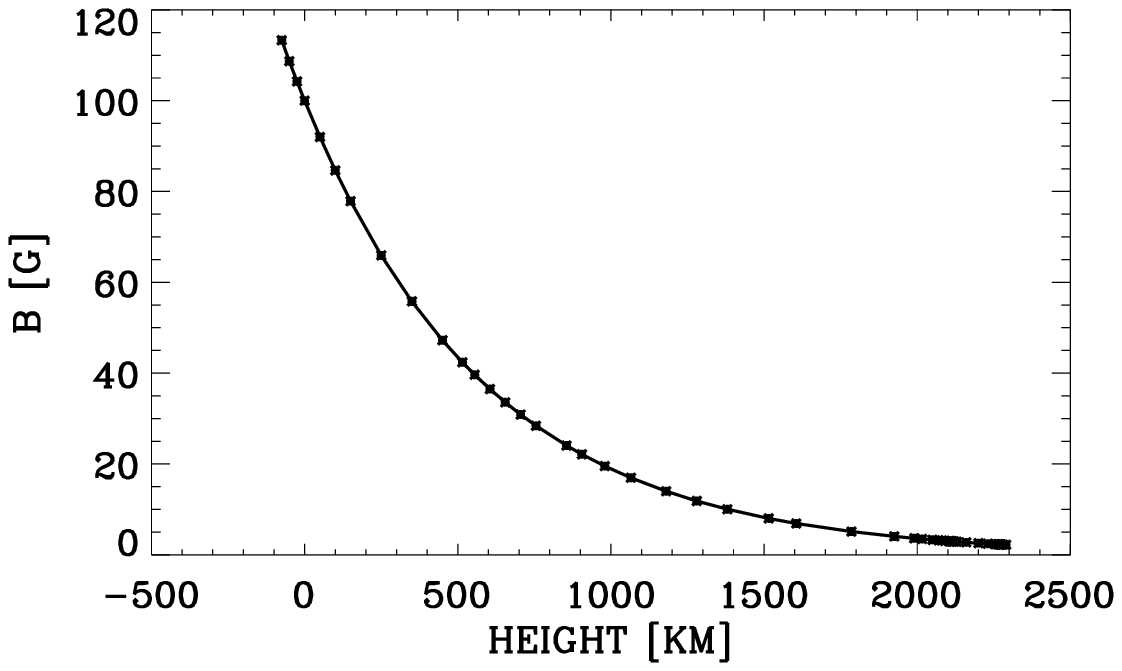}
\includegraphics[width=7cm]{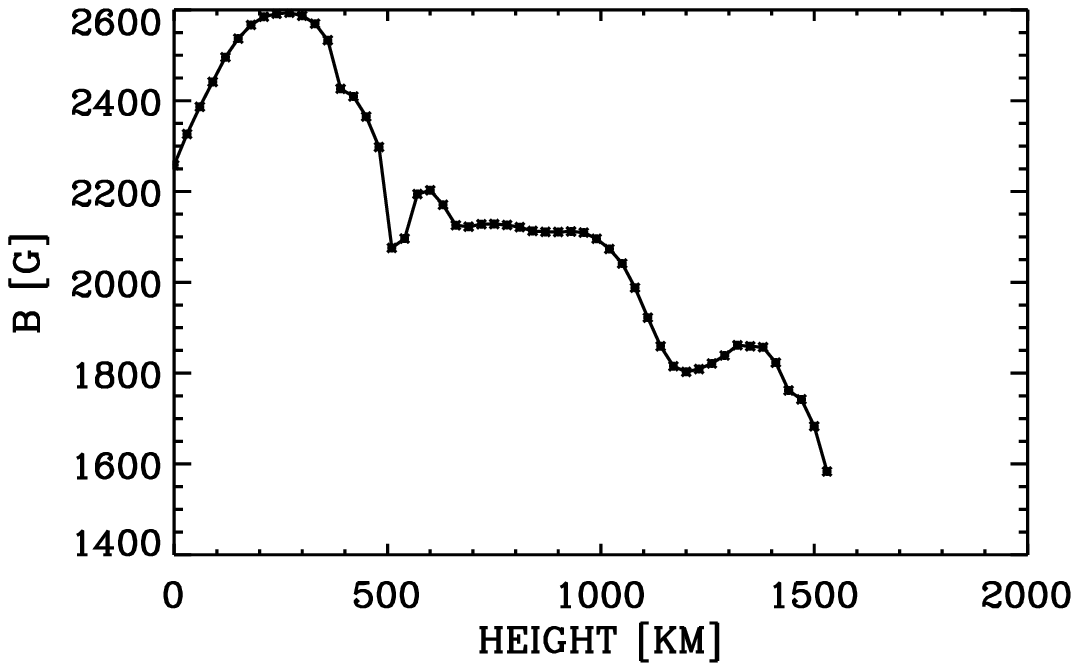}
\caption{{\footnotesize Magnetic field as a function of height assumed for the quiet Sun model (left) and for the sunspot umbra model (right).}
}\label{fig:magfield}
\end{figure*}

\begin{figure*}[t]
\center
\includegraphics[width=7cm]{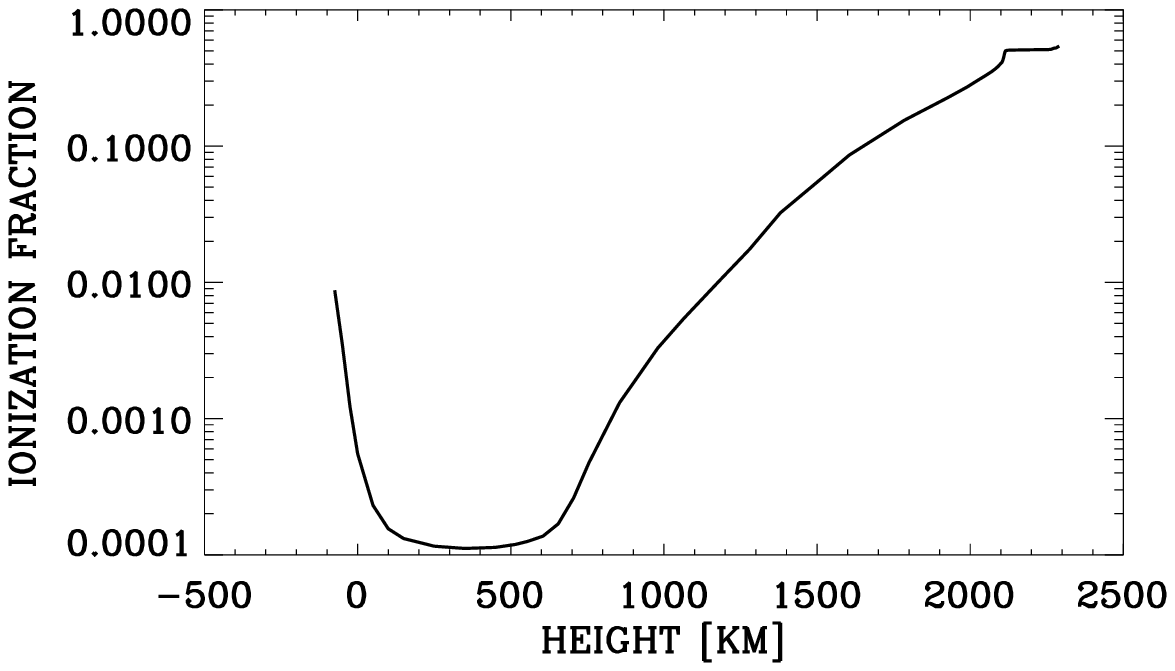}
\includegraphics[width=7cm]{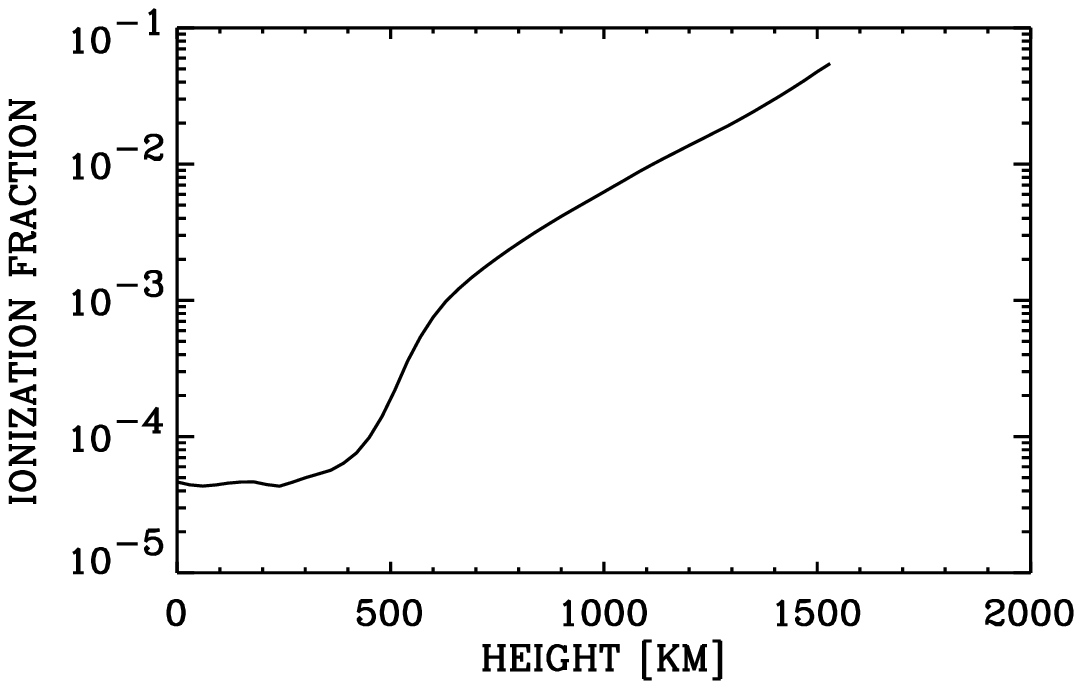}
\caption{{\footnotesize Ionization fraction $\xi_i$ as a function of height in the the VALC model (left) and in the sunspot umbra model (right).}
}\label{fig:ionization}
\end{figure*}
\begin{figure*}[t]
\center
\includegraphics[width=7cm]{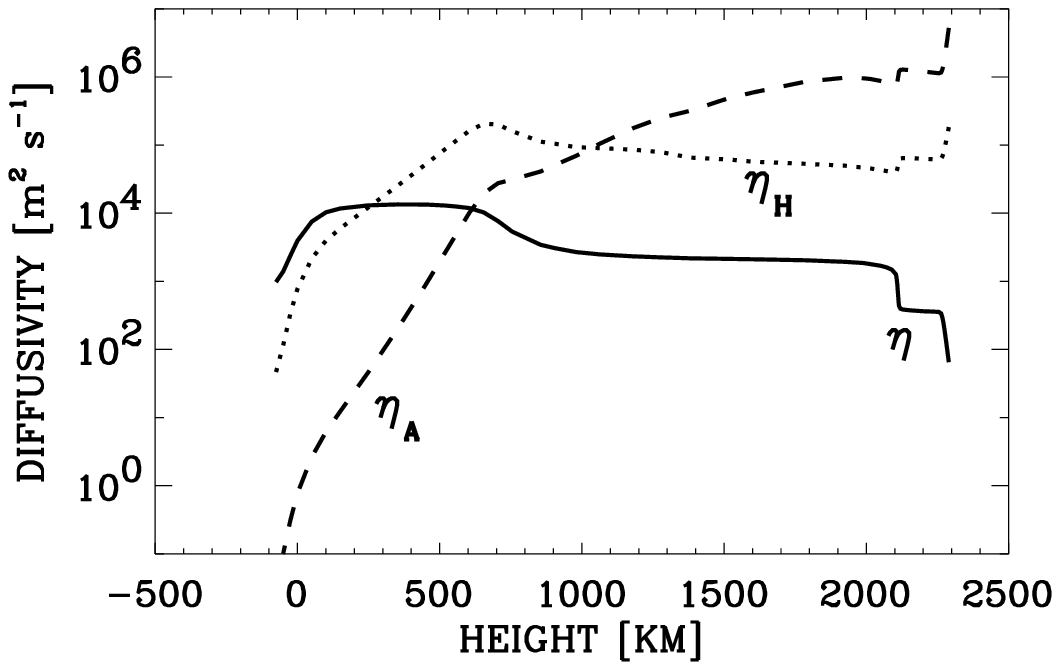}
\includegraphics[width=7cm]{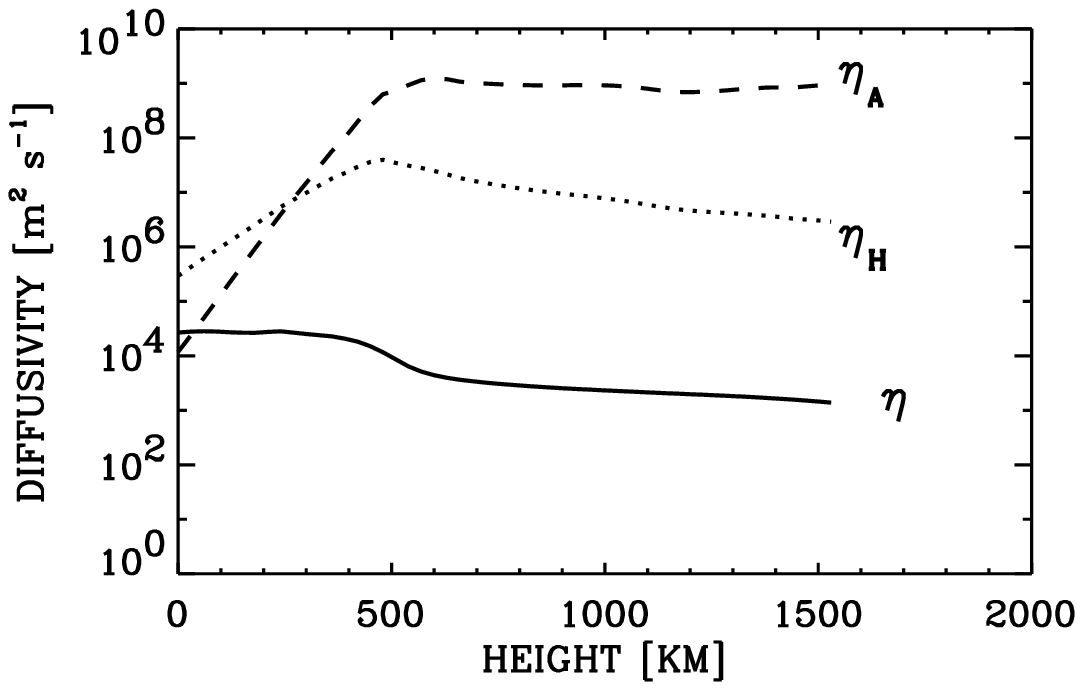}
\caption{{\footnotesize Diffusion parameters from the induction equation for the quiet  Sun model (left) and the sunspot umbra  model (right). Note the difference in scales at the vertical axis.} }\label{fig:etas}
\end{figure*}

\subsection{Ohm's law for single-fluid description}

Using the same strategy as for the two-fluid description, we multiply the individual momentum equations of each species by $q_\alpha/m_{\alpha}$ and add them up. This leads to an equation similar to Eq.~\ref{eq:ohm-bit1}, except that average center of mass velocity $\vec{u}$ appears after the summation, instead of the velocity of charges ($\vec{u}_c$) and a different definition of the drift velocities, $\vec{w}_{\alpha i}$, now measured with respect to the average velocity, $\vec{u}$, rather than with respect to the average velocity of the charges, $\vec{u}_c$:
\begin{eqnarray} \label{eq:ohm-single-bit}
\frac{\partial\vec{J}}{\partial t} + \vec{\nabla}(\vec{J}\otimes \vec{u} + \vec{u}\otimes \vec{J})  + \vec{\nabla} \sum_{\alpha=1}^{N+1}(n_{\alpha}q_{\alpha}\vec{w}_{\alpha}\otimes\vec{w}_{\alpha}) & = & \nonumber \\
\sum_{\alpha=1}^{N+1}\left( \frac{n_\alpha q_\alpha^2}{m_{\alpha}}(\vec{E} + \vec{u}_\alpha \times \vec{B}) - \vec{\nabla}\frac{q_\alpha}{m_{\alpha}}{\bf\hat{p}}_\alpha  + \frac{q_\alpha}{m_{\alpha}} \vec{R}_\alpha \right) & &
\end{eqnarray}
where the summation is done only over the charged components (ions and electrons) since $q_n=0$.

The summation of the Lorentz force term leads to the following
\begin{eqnarray}
&&\sum_{\alpha=1}^{N+1}\frac{n_{\alpha} q_{\alpha}^2}{m_{\alpha}}(\vec{E} + \vec{u}_{\alpha} \times \vec{B})  \approx \frac{e^2 n_e}{m_e}[\vec{E} +\vec{u} \times\vec{B}] + \nonumber \\
&&+ \frac{e^2}{m_e}\sum_{\alpha=1}^{N}n_\alpha \vec{w}_{\alpha}\times\vec{B}  - \frac{e}{m_e}\vec{J}\times\vec{B}
\end{eqnarray}
where we have neglected the term proportional to $m_e/m_{\alpha}$. The summation in the above equation goes only over $N$ ions. We further assume that the drift velocity of individual ions is not different from their average drift velocity, i.e.,
\begin{equation}
\vec{w}_{\alpha} \approx \vec{u}_i - \vec{u}
\end{equation}
with $\vec{u}_i$ being center of mass velocity of ions. Using the definition from Eq.~\ref{eq:totalu}, the relation between the center of mass velocities of charges $\vec{u}_c$, neutrals $\vec{u}_n$, and the average one over the whole fluid $\vec{u}$ can be approximately expressed as:
\begin{equation}
\vec{u} = \frac{\rho_c}{\rho}\vec{u}_c +\frac{\rho_n}{\rho}\vec{u}_n = \xi_c \vec{u}_c + \xi_n\vec{u}_n
\end{equation}
The latter equation provides that $\xi_n + \xi_c = 1$. Combining the two relations above one obtains:
\begin{equation}
\vec{u}_i - \vec{u} = \xi_n(\vec{u}_c - \vec{u}_n) + (\vec{u}_i - \vec{u}_c) \approx  \xi_n\vec{w}
\end{equation}
The center of mass velocity of ions $\vec{u}_i$ and of charges $\vec{u}_c$ (the latter including the contribution from electrons, see Eq.~\ref{eq:uc-alpha}) can be assumed approximately the same if one neglects the electron mass compared to the ion mass. This assumption has already been used in this section, as well as for the derivation of the Ohm's law for the two-fluid formalism. In the equation above we have defined the relative ion-neutral velocity $\vec{w}$ as:
\begin{equation}
\vec{w}=\vec{u}_c - \vec{u}_n \approx \vec{u}_i - \vec{u}_n
\end{equation}

\noindent The summation of the Lorentz force term finally gives:
\begin{eqnarray}
\sum_{\alpha=1}^{N+1}\frac{n_\alpha q_\alpha^2}{m_{\alpha}}(\vec{E} + \vec{u}_\alpha \times \vec{B}) & \approx &  \frac{e^2 n_e}{m_e}[\vec{E} +\vec{u} \times\vec{B}] + \nonumber \\ 
\frac{e^2n_e}{m_e}\xi_n\vec{w}\times\vec{B}  &-& \frac{e}{m_e}\vec{J}\times\vec{B}
\end{eqnarray}

The summation of the pressure terms, and the elastic collision $\vec{R}$ terms, are both the same as in Eq.~\ref{eq:psum-ohm} and Eq.~\ref{eq:rsum-ohm}, correspondingly.

Assuming that currents are stationary in Eq.~\ref{eq:ohm-single-bit}, the partial derivative vanishes. For situations not far from equilibrium, velocities can be assumed to be small and second-order terms including double products can be neglected. Thus, all the terms on the left-hand side of Eq.~\ref{eq:ohm-single-bit} can be removed. The equation for the electric field $\vec{E}$ then reads as:
\begin{eqnarray} \label{ohm_bit1}
\vec{E}^*&=&[\vec{E} + \vec{u}\times{B}] = - \xi_n \vec{w} \times \vec{B}  + \frac{\vec{J}\times \vec{B}}{n_e e}  -\frac{\vec{\nabla}{\bf\hat{p}}_e}{en_e} \nonumber \\
&+& \frac{\rho_e}{(en_e)^2}\left(\sum_{\alpha=1}^N \nu_{ei_\alpha} + \sum_{\beta=1}^N \nu_{en_\beta} \right)\vec{J} \nonumber \\
&-& \frac{\rho_e}{en_e}\left( \sum_{\beta=1}^N\nu_{en_\beta} - \sum_{\alpha=1}^N\sum_{\beta=1}^N\nu_{i_\alpha n_\beta} \right)\vec{w}
\end{eqnarray}

Unlike the two-fluid approach, we can not directly use this equation, because there are still terms depending on $\vec{w}$ and we need further assumptions to proceed from here.

We need the equation for the relative ion-neutral velocity $\vec{w}$ and we obtain it following \citet{Braginskii1965}. We add up the ion-electron and the neutral momentum equations (Eq.~\ref{eq:momentum-alfa-ie-noS} and Eq.~\ref{eq:momentum-alfa-n-noS}), multiplied by $\xi_n$ and $-\xi_c$, correspondingly. The result is:
\begin{eqnarray}
\xi_n\left(\frac{\partial (\rho_c\vec{u}_c)}{\partial t} + \vec{\nabla}(\rho_c\vec{u}_c \otimes \vec{u}_c) \right) &-&\nonumber \\
\xi_c\left(\frac{\partial (\rho_n\vec{u}_n)}{\partial t} + \vec{\nabla}(\rho_n\vec{u}_n \otimes \vec{u}_n) \right) &=& \nonumber \\
\xi_n \left[\vec{J} \times\vec{B} \right] - \vec{G} + \sum_{\beta=1}^N\rho_e\nu_{en_\beta}\frac{\vec{J}}{n_e e} & - & \alpha_n\vec{w}
\end{eqnarray}
where $\vec{G}$ and $\alpha_n$ are given by
\begin{equation}
\vec{G} = \xi_n \vec{\nabla}{\bf\hat{p}}_{ie} - \xi_i \vec{\nabla}  {\bf\hat{p}}_n
\end{equation}
\begin{eqnarray}
\alpha_n &=& \sum_{\beta=1}^N\rho_e\nu_{en_\beta} + \sum_{\alpha=1}^N\sum_{\beta=1}^N\rho_i\nu_{i_\alpha n_\beta}
\end{eqnarray}
The gravity term cancels out since $\vec{g}(\rho_c\xi_n - \xi_c\rho_n)=0$

We can express the total derivative in terms of the drift velocity
\begin{eqnarray}
\xi_n\left(\frac{\partial (\rho_c\vec{u}_c)}{\partial t} + \vec{\nabla}(\rho_c\vec{u}_c \otimes \vec{u}_c) \right) &-& \\
\xi_c\left(\frac{\partial (\rho_n\vec{u}_n)}{\partial t} + \vec{\nabla}(\rho_n\vec{u}_n \otimes \vec{u}_n) \right) &\approx& \rho\xi_n\xi_c\frac{\partial \vec{w}}{\partial t} \nonumber
\end{eqnarray}
\noindent and this term is neglected when compared to the friction terms, the latter assumed to be $\sim \vec{w}/\tau_{\rm col}$, where $\tau_{\rm col} \sim 1/\nu_{\alpha\beta}$ is a typical collisional time scale, see \citet{Diaz+etal2013}. Then the expression for $\vec{w}$ becomes:
\begin{equation}
\label{eq:w_bis}
\vec{w} = \frac{\xi_n}{\alpha_n} \left[\vec{J} \times\vec{B} \right] - \frac{\vec{G}}{\alpha_n} + \sum_{\beta=1}^N\rho_e\nu_{en_\beta}\frac{\vec{J}}{n_e e \alpha_n}
\end{equation}

\noindent Introducing this result in Eq.~\ref{ohm_bit1} and rearranging terms, one can obtain an expression for the electric field $\vec{E}^*$:
\begin{eqnarray}
\label{eq:total_ohm}
\vec{E}^*= c_j \vec{J} & +& c_{jb} [\vec{J} \times \vec{B}]  + c_{jbb} [(\vec{J} \times \vec{B}) \times \vec{B}] + c_{pe} \vec{\nabla}{\bf\hat{p}}_e+ \nonumber  \\
c_{pt} \vec{G} &+& c_{ptb} [\vec{G} \times \vec{B}]
\end{eqnarray}
\noindent where the coefficients of the Ohm's law are given by:
\begin{eqnarray} \label{eq:ohm-coefs1}
\label{c_J}
c_j & = & \frac{1}{(en_e)^2}\left(\sum_{\alpha=1}^N\rho_e\nu_{ei_\alpha} + o\sum_{\beta=1}^N\rho_e\nu_{en_\beta}\right) \approx \frac{\alpha_e}{(en_e)^2}  \nonumber \\
c_{jb} & = & \frac{1}{en_e}\left(1 - 2\xi_n\epsilon_1 + \xi_n\epsilon_2 \right) \nonumber \\
c_{jbb} & = & -\frac{\xi_n^2}{\alpha_n} \nonumber \\
c_{pe} & = & -\frac{1}{en_e} \nonumber \\
c_{pt} & = & \frac{1}{e n_e} \left(\epsilon_1 - \epsilon_2 \right) \nonumber \\
c_{ptb} & = & \frac{\xi_n}{\alpha_n}
\end{eqnarray}
\noindent In these equations, we have defined the additional parameters:
\begin{equation}
\alpha_e=\sum_{\alpha=1}^N\rho_e\nu_{ei_\alpha} + \sum_{\beta=1}^N\rho_e\nu_{en_\beta}
\end{equation}
\begin{eqnarray}
\epsilon_1&=&\sum_{\beta=1}^N\rho_e\nu_{en_\beta}/\alpha_n \ll 1 \\
\epsilon_2&=&\sum_{\alpha=1}^N\sum_{\beta=1}^N\rho_e\nu_{i_\alpha n_\beta}/\alpha_n \ll 1 \\
o&=&\sum_{\alpha=1}^N\sum_{\beta=1}^N\rho_{\alpha i}\nu_{i_\alpha n_\beta}/\alpha_n \approx 1
\end{eqnarray}
according to the definition of the collisional frequencies Eq.~\ref{eq:nus-approximated}.

The expression for the coefficient $c_{jbb}$ in Eq. \ref{eq:ohm-coefs1} can be compared with those given in other studies that include the contribution of neutral helium by \citet{Zaqarashvili+etal2013} and \citet{Soler2010}. Our expression for this coefficient is more compact than given in \citet{Zaqarashvili+etal2013} and is more similar to  \citet{Soler2010}.  \citet{Zaqarashvili+etal2013} showed that the expression by \citet{Soler2010} does not have a general validity and is only valid for weakly ionized medium, since equal velocities of neutral hydrogen and helium were assumed for its derivation. In our case, we have not made any assumptions about the ionization degree of the plasma, but we supposed that the terms $\sum_\alpha n_{\alpha i}\vec{w}_{\alpha i}$ and $\sum_\beta n_{\beta n}\vec{w}_{\beta n}$ representing the sum of the individual drift velocities of ion and neutral species with respect to the average ion and neutral velocity are small. Therefore, our expression does not suffer the same restrictions as the one by \citet{Soler2010}.  


\begin{figure*}[t]
\center
\includegraphics[width=7cm]{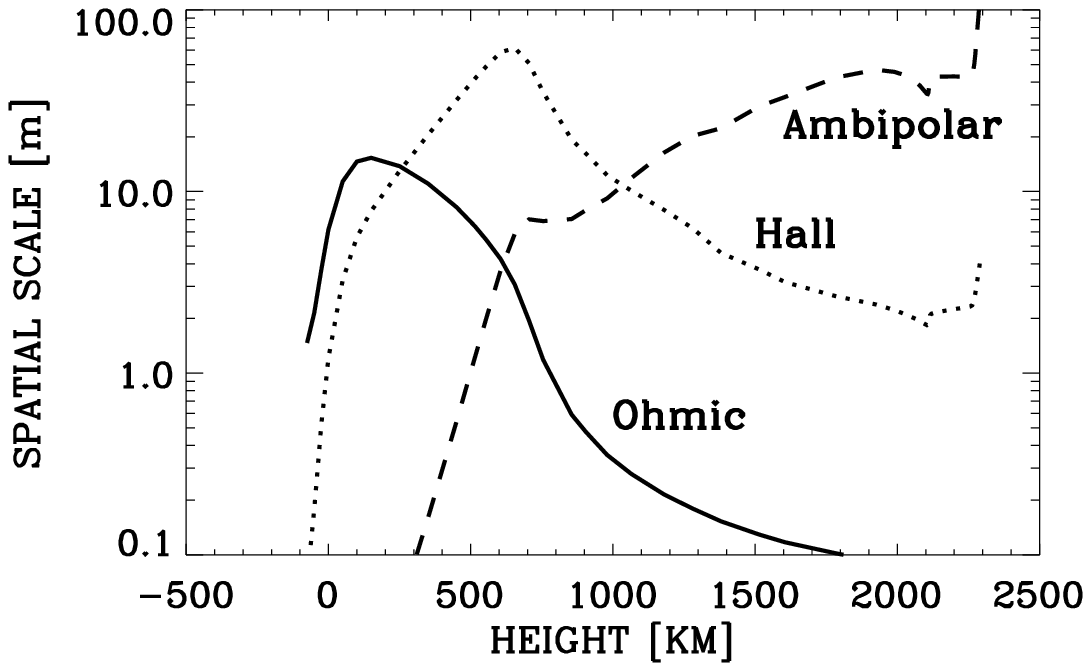}
\includegraphics[width=7cm]{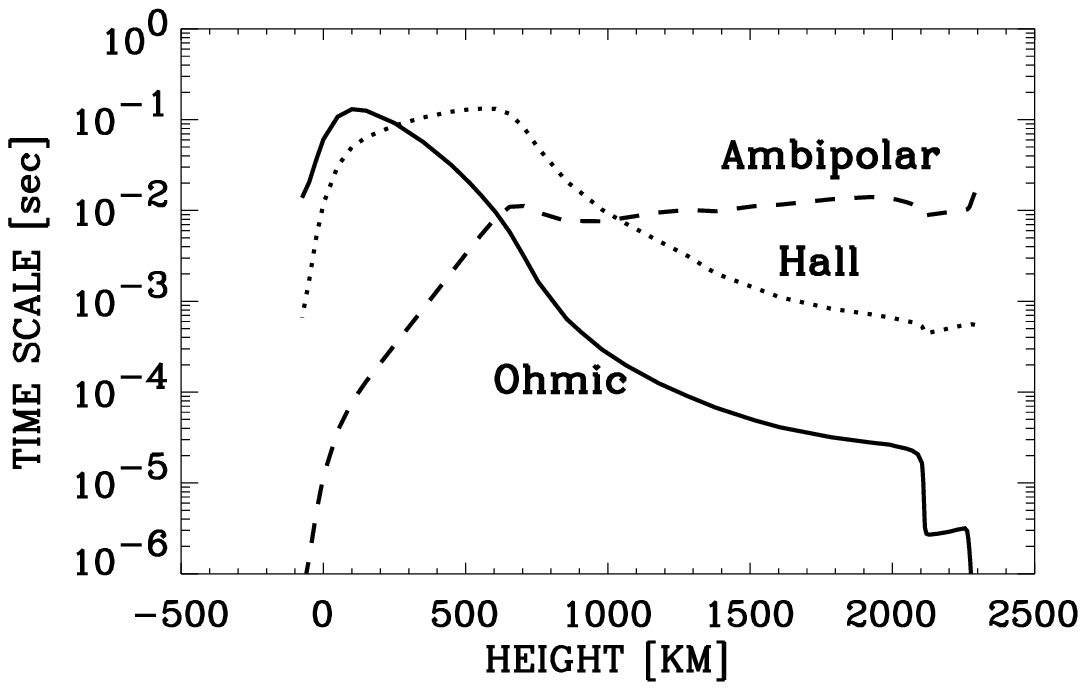}
\caption{{\footnotesize Spatial and temporal scales defining the range when the Ohmic, Hall and Ambipolar terms become important relative to the convection term in the induction equation, for the quiet Sun model field.}}\label{fig:scales2}
\end{figure*}

\begin{figure*}[t]
\center
\includegraphics[width=7cm]{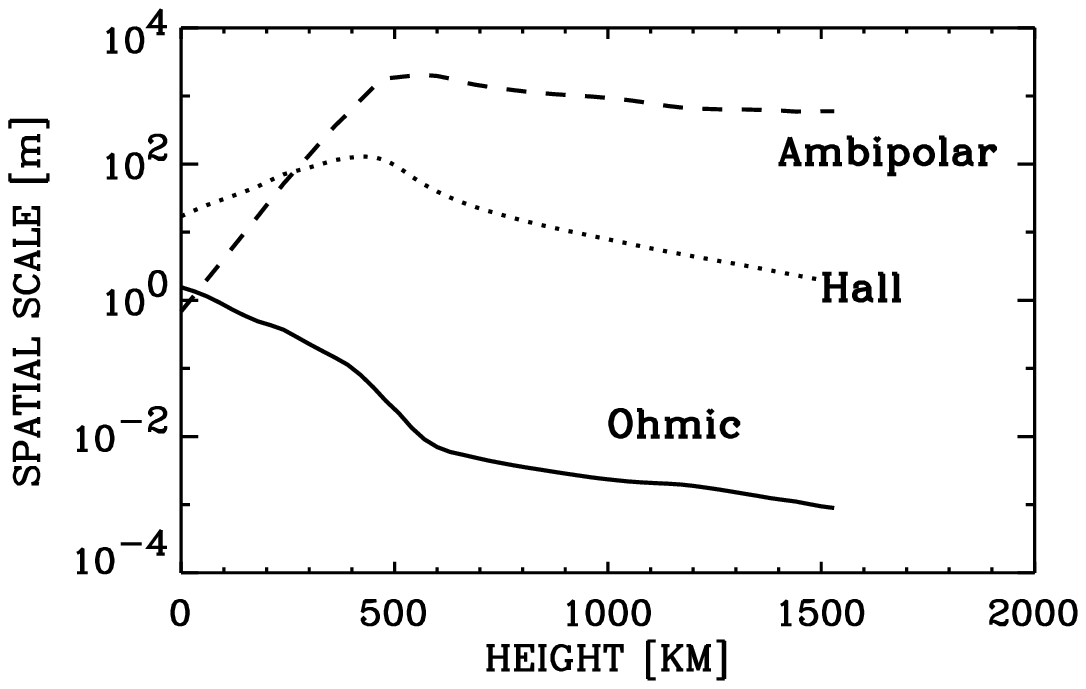}
\includegraphics[width=7cm]{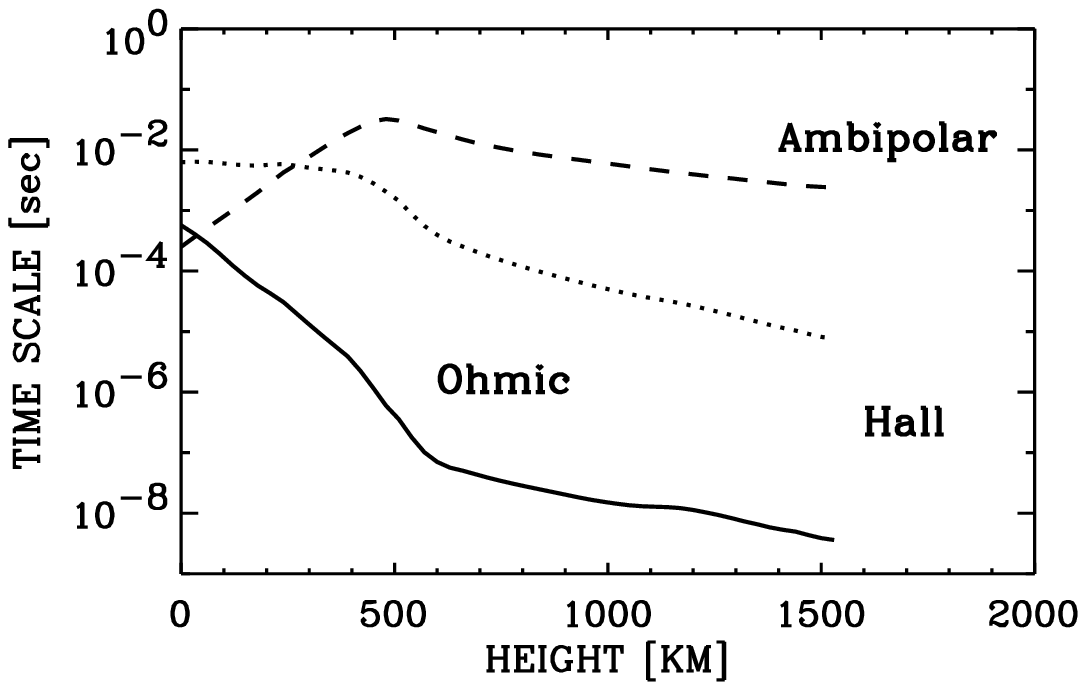}
\caption{{\footnotesize Spatial and temporal scales defining the range when the Ohmic, Hall and Ambipolar terms become important relative to the convection term in the induction equation, for the sunspot umbra model.}
}\label{fig:scales2_hsocas}
\end{figure*}
\begin{figure*}[t]
\center
\includegraphics[width=16cm]{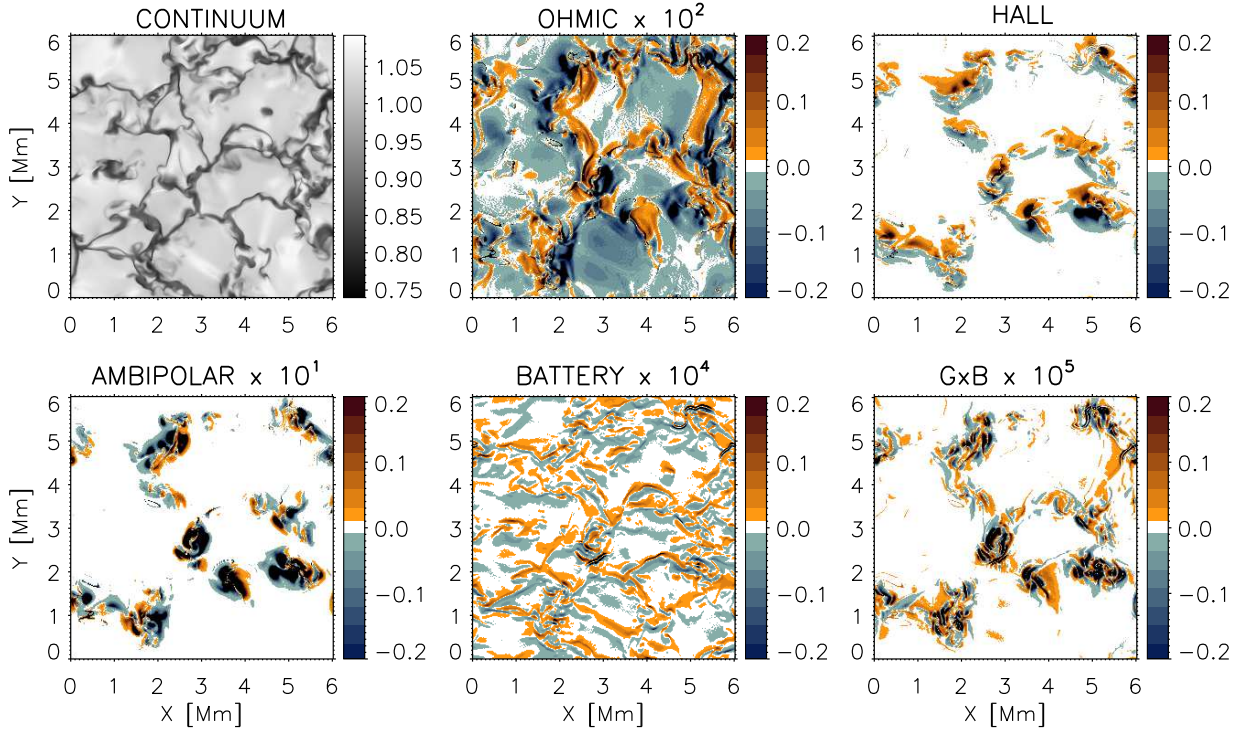}
\caption{{\footnotesize Spatial distribution of the terms of the Ohm's law in a snapshot of magneto-convection with the average magnetic field strength of 180 G. From left to right, from top to bottom, the panes give the continuum intensity, and the $y$ components of the following terms: the Ohmic term $\eta\mu J_y$, Hall term $\eta_H\mu(\vec{J}\times\vec{B})_y/|B|$, Ambipolar term $\eta_A\mu([\vec{J}\times\vec{B}]\times\vec{B})_y/|B|^2$, battery term $\eta_p\mu dP_e/dy /|B|$, and the pressure term $c_{ptb}(\vec{G}\times\vec{B})_y$. The units are SI units ($mlq^{-1}t^{-2}$). The different terms are multiplied by a factor (when necessary) so that the scale is the same in all the panels. The Hall term is taken as a reference. The larger is the multiplying factor, the smaller is the term. The quantities are given at height of 600 above the photosphere.} } \label{fig:mconv}
\end{figure*}

\begin{figure*}[t]
\center
\includegraphics[width=16cm]{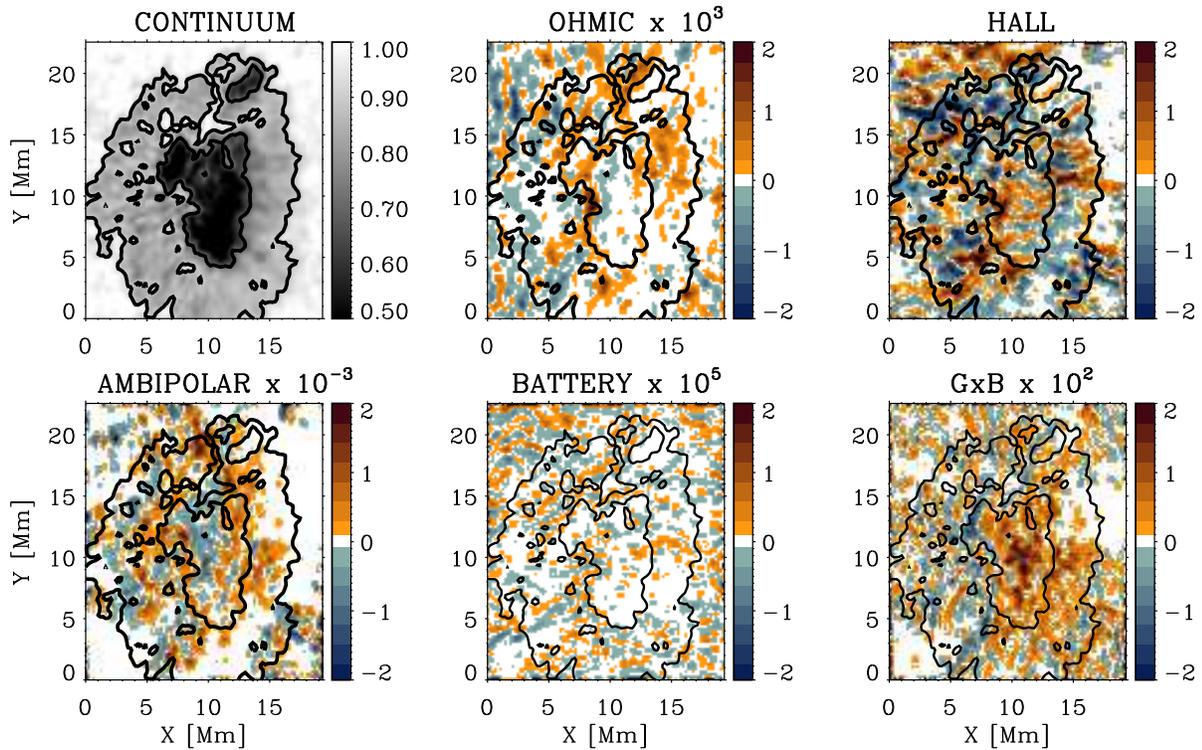}
\caption{{\footnotesize Spatial distribution of the terms of the Ohm's law in a 3D semi-empirical sunspot umbra model. From left to right, from top to bottom, the panes give the continuum intensity, and the $y$ components of the following terms: the Ohmic term $\eta\mu J_y$, Hall term $\eta_H\mu(\vec{J}\times\vec{B})_y/|B|$, Ambipolar term $\eta_A\mu([\vec{J}\times\vec{B}]\times\vec{B})_y/|B|^2$, battery term $\eta_p\mu dP_e/dy /|B|$, and the pressure term $c_{ptb}(\vec{G}\times\vec{B})_y$. The units are SI units ($mlq^{-1}t^{-2}$). The different terms are multiplied by a factor (when necessary) so that the scale is the same in all the panels. The Hall term is taken as a reference. The larger is the multiplying factor, the smaller is the term. The quantities are given at height of 1500 in the chromosphere.} } \label{fig:spot}
\end{figure*}

\subsection{Total induction equation}

To get the induction equation, we use Ohm's law, Eq.~\ref{eq:total_ohm}, and Faraday's law and Ampere's law neglecting Maxwell's displacement current:
\begin{equation}
\frac{\partial\vec{B}}{\partial t} = -\vec{\nabla} \times \vec{E}; \,\,\, \vec{J} = \frac{1}{\mu}\vec{\nabla} \times \vec{B}
\end{equation}

\noindent The generalized induction equation can be rewritten in terms of current density $\vec{J}$ as:
\begin{eqnarray} \label{eq:induction-final}
\frac{\partial\vec{B}}{\partial t}  =  \vec{\nabla}\times \left[(\vec{u}\times\vec{B}) - \eta\mu\vec{J} - \frac{\eta_H\mu}{|B|}[\vec{J}\times\vec{B}] + \right. \nonumber \\ \left. + \frac{\eta_A\mu}{|B|^2}[[\vec{J}\times\vec{B}]\times\vec{B}]  + \frac{\eta_p\mu}{|B|}\vec{\nabla}{\bf\hat{p}}_e  - \right. \nonumber \\ \left. - c_{pt}\vec{G} - c_{ptb}[\vec{G}\times\vec{B}] \right]
\end{eqnarray}
\noindent where we have defined the diffusivity coefficients, making use of Eqs.~\ref{eq:ohm-coefs1}.
\begin{eqnarray} \label{eq:eta}
\eta & = & c_j/\mu= \frac{\alpha_e}{(en_e)^2\mu}  \\
\eta_H & = & c_{jb}/\mu |B| =  \frac{|B|}{en_e\mu} \\
\eta_A & = & -c_{jbb}/\mu |B|^2 = \frac{\xi_n^2}{\alpha_n\mu}|B|^2 \\
\eta_p & = & -c_{pe}/\mu |B| = \frac{|B|}{en_e\mu}
\end{eqnarray}

\noindent In the induction equation, the terms on the right hand side are: convective term, Ohmic term, Hall term, ambipolar term, Bierman battery term, terms depending on the pressure gradients distribution and gravity terms. The vector $\vec{G}$ becomes small if the neutral fraction is small, i.e. in the weakly ionized plasma.

Note that in the above definitions of diffusion coefficient we formally scaled $\eta_p$ with $|B|/\mu$ to have the same dimensions as other diffusion coefficients $\eta$, $\eta_H$, $\eta_A$ ($l^2/t$, i.e. $m^2s^{-1}$ in the SI units). The expression of the coefficient multiplying the Biermann battery pressure term $\vec{\nabla}{\bf\hat{p}}_e$ is exactly the same as of the Hall term $[\vec{J}\times\vec{B}]$. The relative importance of the both terms is thus determined by the strength of the current density $J$, magnetic field $B$ and the value of the electron pressure (that may be small for a weakly ionized plasma), as well as the amplitude of spatial variations of these quantities.

Another often used parameter is Cowling conductivity \citep{Khodachenko2004, Khodachenko2006, Leake+Arber2006, Forteza2007, Arber2007, Arber2009, Sakai+Smith2009}. It can be introduced combining the Ohm and Ambipolar terms:
\begin{equation}
\eta_A\mu=\frac{\xi_n^2|B|^2}{\alpha_n}=\mu\eta_c-\mu\eta=1/\sigma_c - 1/\sigma
\end{equation}
\begin{equation} \label{eq:cowling}\sigma_c=\frac{\sigma}{1 + \sigma\frac{\xi_n^2|B|^2}{\alpha_n}}=\frac{\sigma}{1 + \sigma\eta_A\mu}
\end{equation}

\noindent By introducing this conductivity, the ambipolar term, together with the Ohmic term can be rewritten as:
\begin{equation}
\eta\mu\vec{J}  - \eta_A \mu \frac{( \vec{J} \times \vec{B} ) \times \vec{B}}{|B|^2} =\eta\mu\vec{J}_{||} + \eta_c \mu\vec{J}_{\bot}
\end{equation}
so that the Cowling conductivity is responsible for the perpendicular currents to the magnetic field. The Cowling conductivity depends on the magnetic field and on the ionization fraction. If there are no neutrals ($\xi_n=0$), both Ohmic and Cowling conductivity are the same.

\section{Ohm's law in the solar atmosphere}

To estimate the relative importance of the Ohmic, Hall, Ambipolar and other terms in the induction equation, one has to compare the values of $\eta$, $\eta_H$, $\eta_A$, $\eta_p$, etc. for the typical conditions of the solar atmosphere.

We compare two atmospheric models. In the first model we take thermodynamic parameters from the VALC model atmosphere \citep{valc}, and the magnetic field is assumed to depend on height as $B=100\exp{(-z/600)}$ G, roughly representing a quiet solar atmosphere outside active regions. The second model is extracted from the non-LTE inversion of the IR Ca triplet by \citet{Socas-Navarro2005} and represents a sunspot umbra. These two models give the two extreme cases of the conditions that can be found in the solar photosphere and chromosphere.

Figures \ref{fig:magfield} and \ref{fig:ionization} show the magnetic field stratification and the ionization fraction $\xi_i$ in the both models. The magnetic field reaches the maximum of 2.5 kG in the sunspot model, but only 100 G in the quiet Sun model. The ionization fraction is always below 1. In the quiet Sun the minimum of $\sim 10^{-4}$ is reached in the photosphere. In the sunspot model, the ionization fraction is even lower, as expected for the cooler temperatures appropriate for the umbra.

The diffusion coefficients for the both models are shown in Figure \ref{fig:etas}. This figure gives the coefficients $\eta$, $\eta_H$ (equal to $\eta_p$ in our definition) and $\eta_A$. The comparison of the diffusion coefficients for the quiet model (left) shows that the Hall term becomes more important at heights above 200, where the electrons decouple from the rest of the plasma, according to Fig.~\ref{fig:scales}. At this height the cyclotron frequency of electrons becomes larger than their collisional frequency. The Ambipolar term becomes more important above 1000 km when the ions decouple from the rest, according to Fig.~\ref{fig:scales}.

As for the sunspot model, the whole stratification seems to be shifted downward in height (Willson depression). The Ohmic diffusivity is of the same  order of magnitude as in the VALC model. However both Hall and Ambipolar terms are two orders of magnitude larger due to the larger magnetic field. The Ambipolar term dominates over almost all height range starting from 300 km in the photosphere.

To define the typical temporal and spatial scales where the Ohmic, Hall and Ambipolar terms become important, compared to the convective term $\vec{u}\times\vec{B}$, we have to relate them between each other. If $u$ is of the order of Alfv\'en speed $v_A=B/\sqrt{\mu\rho}$, then the spatial scale is given by:
\begin{eqnarray}
L & = & \frac{\eta}{v_A}
\end{eqnarray}
The temporal scale is $\nu=v_A/L$ making it as:
\begin{equation}
\tau=\frac{1}{\nu}=\frac{\eta}{v_A^2}
\end{equation}
In the above equations $\eta$ stays for either Ohmic, Hall or Ambipolar diffusivity coefficient. 

The scales go with $B$ as:
\begin{eqnarray}
L_{\rm ohm} & \sim &  B^{-1} \\
\tau_{\rm ohm} & \sim &  B^{-2} \\
L_H & = &  {\rm const}(B) \\
\tau_H & \sim & B^{-1} \\
L_A & \sim &  B  \\
\tau_A & = &  {\rm const}(B)
\end{eqnarray}
so the importance of the Ambipolar diffusion should increase in the strong-field structures, while the importance of the Hall effect should increase in the weak-field structures.

Another way of evaluating the importance of these terms is by comparing their scales with typical frequencies of the atmosphere, such as cyclotron frequencies and collisional frequencies. Using the expressions for the $\eta$ coefficients above, one readily obtains
\begin{eqnarray}
\tau_{\rm ohm} & \sim & \frac{\sum_{\alpha=1}^N\nu_{ei_\alpha} + \sum_{\beta=1}^N\nu_{en_\beta}}{\omega_{ci}\omega_{ce}} \\
\tau_H & \sim & \frac{1}{\omega_{ci}} \\
\tau_A & \sim & \frac{\xi_n^2}{\sum_{\alpha=1}^N\sum_{\beta=1}^N\nu_{i_\alpha n_\beta}} \,.
\end{eqnarray}
Therefore, the Hall term becomes important for time scales close to the ion cyclotron period, the Ambipolar term becomes important on scales proportional to ion-neutral collision time \citep[see][]{Zaqarashvili2011} and the Ohmic term is important on scales given by the ratio of the electron collisional frequency to the product of  the ion and electron cyclotron  frequencies.

The typical scales obtained from the above estimation are given in Fig.~\ref{fig:scales2} for the quiet model and in Fig.~\ref{fig:scales2_hsocas} for the sunspot umbra model. In the quiet model the largest spatial scale for the Hall effect reaches 100 m at 600 km; the largest temporal scale is 0.1 sec, also for the Hall effect in the upper photosphere. For the Ambipolar term, the largest scales are at the upper chromosphere, reaching the same values.

In the sunspot umbra model, the Ohmic term only important at spatial scales below 1 m, the Hall term in the photosphere reaches the scales of 100 m, while the Ambipolar term dominates at scales reaching few 1000 m. The temporal scales in the umbral model do not exceed 0.1 sec for all the terms. The spatial and temporal scales calculated for the sunspot umbra model agree with those given in \citet{Pandey+Wardle2008}.

It is also interesting to check the relative importance of the other terms in the Ohm's law compared to the Ohmic, Hall and Ambipolar terms. For that we make use of the three dimensional solar model atmospheres. We use a snapshot of solar magneto-convection calculated by \citet{Vogler} as a representative of the quiet Sun plage region. The snapshot has horizonal spatial resolution of about 20 km, a mostly unipolar magnetic field distributed on granular scales with the average strength of 180 G. The maximum height reached in this magneto-convection model is about 650 km above the photosphere. The second model is the same sunspot model from \citet{Socas-Navarro2005} as we used before, but now all three dimensions are used instead of an average representative over the umbral region. The horizontal pixel resolution of this model is about 200 km and the vertical resolution is 30 km. The maximum height reached is about 2000 km in the chromosphere.

The Ohm's law gives a vector of the current density $J$. As a matter of an example we display only the horizontal $y$ component of the current $E_y$ at a given height. Figures \ref{fig:mconv} and \ref{fig:spot} show the the result of the calculation. In the quiet region (Fig.~\ref{fig:mconv}) Ohmic, Hall and Ambipolar are the dominating terms. The latter two are largest at the borders of granules where the gradients are the strongest. The amplitude of the variations of these terms over the surface is several orders of magnitude, strongly depending on the magnetic field and the ionization fraction (i.e. temperature). The battery and the pressure term (last two panels) are, on average $4 - 5$ orders of magnitude smaller. However, we should take this sample calculation with care because the magneto-convection simulations were computed without taking into account the non-ideal terms of the Ohm's law and because the spatial resolution of this snapshot is not enough to resolve the scales where these terms can be important (see Fig.~\ref{fig:scales2}).

In the sunspot model (Fig.~\ref{fig:spot}) the relative contribution of the different terms changes. The ambipolar term dominates by 3 orders of magnitude over all other terms. The next largest term is the Hall term. The Ohmic and the pressure term $\vec{G}\times \vec{B}$ are of the same order of magnitude, being $2-3$ orders of magnitude smaller than the Hall term. The battery term is not important, however. Once again, this calculation has to be taken with care because of the relatively low spatial resolution in the derivatives (200 km) and all other approximation assumed when doing the inversion of the IR Ca triplet profiles \citep[see][]{Socas-Navarro2005}. The spatial variation of the terms also shows hints for dependence of the magnetic and temperature structure of the observed sunspot, though not so much clear as in the numerically computer magneto-convection model with better resolution.

\section{Conclusions}

In this paper we have provided a self-consistent derivation of equations for the description of multi-component, multi-specie solar plasma composed of an arbitrary number of ionized and neutral atoms. Starting from the microscopic description of the  atom interactions at different ionization-excitation states we have derived the macroscopic equations of conservation of mass, momentum and energy, including the interaction of the matter and radiation field. Without providing all the details of radiative interactions (that depend on particular transition in particular atom), we have given several examples for the formulation for different collisional source terms in the conservation equations for the case of radiative excitation process. 

Conservation equations for multiple micro-states were summed out to provide a two-fluid and a single-fluid formalism, in order to reduce the number of variables and necessary to make the system of equations solvable on practice. Unlike other works, our equations cover the case of an arbitrary number of neutrals and  singly-ionized ions, and not just hydrogen or hydrogen-helium plasma. The latter is important for the self-consistent description of solar plasma where most neutral atoms come from hydrogen,  that is essentially not ionized in the photosphere, but most electrons come from other atoms that are ionized even at the relatively low photospheric temperatures. Therefore, in our approach, the ionization fraction can be calculated self-consistently. 

In the two-fluid formalism we write separately equations for an average neutral component and an average ionized component, assuming that the difference in velocity and other parameters between these two components is larger than between the different ions and neutrals of different kind themselves. This assumption is reasonable because the forces acting on neutral and ionized particles are different. Nevertheless, due do their different masses, the force acting on different neutrals themselves can also be different. If the collisional coupling is weak enough, different neutral components may decouple and should be considered apart. Our formalism can be easily extended to cover this situation. It can also be extended, if necessary, to the case of multiple ionizations. However, in the regions of interest of the solar atmosphere  where the concentrations of neutrals is the largest, multiply ionized atoms are not so frequent.

Finally, we have derived Ohm's law for the multi-component plasma, both for the two-fluid and single-fluid cases. We have evaluated the magnitude of the different terms in the Ohm's law (as Ohmic, Hall, Ambipolar and battery terms) for several representative models of the solar atmosphere, covering quiet and active regions. In the quiet regions, Hall and Ambipolar terms dominate the electric current, being largest at the borders of inter granular lanes with strongest gradients of all parameters. In the sunspot model, the Ambipolar term dominates by 3 orders of magnitude over the next in magnitude Hall term. These results can be taken as a guidance for the future application of non-ideal MHD modeling in different solar situations.

\begin{acknowledgements}
This work is partially supported by the Spanish Ministry of Science through projects AYA2010-18029 and AYA2011-24808. This work contributes to the deliverables identified in FP7 European Research Council grant agreement 277829, ``Magnetic connectivity through the Solar Partially Ionized Atmosphere''.
\end{acknowledgements}

\appendix

\section{Definitions for macroscopic equations}
\label{app:pq}

\begin{itemize}

\item Individual pressure tensors:
\begin{eqnarray} \label{eq:app1-p}
{\bf\hat{p}}_{\alpha \mss{IE}} &=&\rho_{\alpha \mss{IE}}\langle \vec{c}_{\alpha \mss{I}} \otimes \vec{c}_{\alpha \mss{I}}\rangle \nonumber \\
{\bf\hat{p}}_{\alpha \mss{I}}&=&\sum_E{\bf\hat{p}}_{\alpha \mss{IE}}=\rho_{\alpha \mss{I}}\langle \vec{c}_{\alpha \mss{I}} \otimes \vec{c}_{\alpha \mss{I}} \rangle  \nonumber \\
{\bf\hat{p}}_e&=&\rho_e\langle \vec{c}_e \otimes \vec{c}_e \rangle
\end{eqnarray}
\item Individual heat flow vectors:
\begin{eqnarray} \label{eq:app1-q}
\vec{q}_{\alpha \mss{IE}}& &= \frac{1}{2}\rho_{\alpha \mss{IE}}\langle c_{\alpha \mss{I}}^2\vec{c}_{\alpha \mss{I}} \rangle \nonumber  \\
\vec{q}_{\alpha \mss{I}} &=& \sum_E \vec{q}_{\alpha \mss{IE}} = \frac{1}{2}\rho_{\alpha \mss{I}}\langle c_{\alpha \mss{I}}^2\vec{c}_{\alpha \mss{I}} \rangle \nonumber  \\
\vec{q}_e &=& \frac{1}{2}\rho_e\langle c_e^2\vec{c}_e \rangle
\end{eqnarray}
\item Total potential energy of ionization level $\mathsc{I}$ of a specie $\alpha$ 
\begin{equation} \label{eq:chi-ai}
\chi_{\alpha \mss{I}}=\sum_{\mss{E}}n_{\alpha \mss{IE}}E_{\alpha \mss{IE}}
\end{equation}
\item Potential excitation-ionization energy source term
\begin{equation} \label{eq:Phi-ai}
\Phi_{\alpha \mss{I}} = \sum_{\mss{E}}E_{\alpha \mss{IE} }S_{\alpha \mss{IE} }/m_{\alpha \mss{I}}
\end{equation}
\end{itemize}
where the subscript $\mathsc{I}=1$ for ions and $0$ for neutrals (also denoted by subscripts ``i'' and ``n'').

\section{Properties of $\vec{R}$ and $S$ terms}
\label{app:SR}

\noindent The total inelastic neutral, ion and electron collisional terms, considering only first ionization, are defined as follows:
\begin{eqnarray} \label{eq:Sie-alpha}
S_n&=&\sum_\alpha m_{\alpha 0}  \sum_{\mss{E}}\sum_{\mss{E'}}(n_{\alpha \mss{1E'}}P_{\alpha \mss{1E' 0E}} - n_{\alpha \mss{0E}}P_{\alpha \mss{0E 1E'}}) \nonumber \\
S_i&=&\sum_\alpha m_{\alpha 1} \sum_{\mss{E}}\sum_{\mss{E'}}(n_{\alpha \mss{0E'}}P_{\alpha \mss{0E' 1E}} - n_{\alpha \mss{1E}}P_{\alpha \mss{1E 0E'}}) \nonumber \\
S_e&=&m_e\sum_\alpha\sum_{\mss{E}}\sum_{\mss{E'}}(n_{\alpha \mss{0E'}}P_{\alpha \mss{0E' 1E}} - n_{\alpha \mss{1E}}P_{\alpha \mss{1E 0E'}}) 
\end{eqnarray}

We use subindices $n$ and $i$ to refer to the source terms for neutrals ($\mathsc{I}=0$) and singly-ionized ions ($\mathsc{I}=1$). Note that $S_n+S_i + S_e=0$. i.e. the total mass is conserved.

The elastic collisional term between the two particles of two different micro-states $\{\alpha \mathsc{IE}\}$ and $\{\beta\mathsc{I'E'}\}$ is defined as
\begin{equation}
\vec{R}_{\alpha \mss{IE} ; \beta\mss{I'E'}}  = - \rho_{\alpha \mss{IE}}\nu_{\alpha \mss{IE} ; \beta\mss{I'E'}}(\vec{u}_{\alpha \mss{IE}} - \vec{u}_{\beta\mss{I'E'}})
\end{equation}
so the collisions between the particles of the kind $\{\alpha \mathsc{IE}\}$ with all other particles are given by
\begin{equation}
\vec{R}_{\alpha \mss{IE}} = -\rho_{\alpha \mss{IE}}\sum_{\beta\mss{I'E'}}\nu_{\alpha \mss{IE} ; \beta\mss{I'E'}}(\vec{u}_{\alpha \mss{IE}} - \vec{u}_{\beta\mss{I'E'}}) 
\end{equation}

\noindent The total collisional term, summed over all excitation states is trivially obtained, since we assume that neither velocities nor collisional frequencies depend on the excitation states:
\begin{eqnarray} \label{eq:r-alpha-i}
\vec{R}_{\alpha \mss{I}} = -\sum_{\mss{E}}\rho_{\alpha \mss{IE}}\sum_{\beta\mss{I'E'}}\nu_{\alpha \mss{IE} ; \beta\mss{I'E'}}(\vec{u}_{\alpha \mss{IE}} - \vec{u}_{\beta\mss{I'E'}}) = \nonumber \\
-\rho_{\alpha \mss{I}}\sum_{\beta\mss{I'}}\nu_{\alpha \mss{I} ; \beta\mss{I'}}(\vec{u}_{\alpha \mss{I}} - \vec{u}_{\beta\mss{I'}})
\end{eqnarray}

Particularizing to the case of singly-ionized ions, we change to a standard nomenclature, and denote singly-ionized ions of a specie $\alpha$ by sub-index $\alpha_i$, and neutrals of a specie $\beta$ by $\beta_n$, the collisional momentum transfer between ions of sort $\alpha$ and the rest of the particles is:
\begin{eqnarray}
\vec{R}_{\alpha i}&=&-\rho_{\alpha i}\nu_{i_\alpha e}(\vec{u}_{\alpha i} - \vec{u}_e) - \rho_{\alpha i}\sum_{\beta}\nu_{i_\alpha n_\beta}(\vec{u}_{\alpha_i} - \vec{u}_{\beta n})\nonumber \\
&-& \rho_{\alpha i}\sum_{\beta}\nu_{i_\alpha i_\beta}(\vec{u}_{\alpha_i} - \vec{u}_{\beta i})
\end{eqnarray}
where the electrons are denoted by sub-index $e$.

Collisional term between all the ions with the rest of the particles is the sum of the last expression over $\alpha$
\begin{eqnarray}  \label{eq:R_i}
\vec{R}_i=\sum_{\alpha}\vec{R}_{\alpha i} =  &-&\sum_{\alpha}\rho_{\alpha i}\nu_{i_\alpha e}(\vec{u}_{\alpha i}- \vec{u}_e) \nonumber \\
&-& \sum_{\alpha}\rho_{\alpha i}\sum_{\beta}\nu_{i_\alpha n_\beta}(\vec{u}_{\alpha_i} - \vec{u}_{\beta n})
\end{eqnarray}
where the ion-ion collisions add up to zero.

Collisional term between neutrals of sort $\beta$ and the rest of the particles is:
\begin{eqnarray}
\vec{R}_{\beta n}&=&-\rho_{\beta n}\nu_{n_\beta e}(\vec{u}_{\beta n} - \vec{u}_e) - \rho_{\beta n}\sum_{\alpha}\nu_{n_\beta i_\alpha}(\vec{u}_{\beta n} - \vec{u}_{\alpha i}) \nonumber \\
&-& \rho_{\beta n}\sum_{\alpha}\nu_{n_\beta n_\alpha}(\vec{u}_{\beta n} - \vec{u}_{\alpha n})
\end{eqnarray}

Collisional term between all the neutrals with the rest of the particles is:
\begin{eqnarray}   \label{eq:R_n}
\vec{R}_n&=&\sum_{\beta}\vec{R}_{\beta n} =  - \sum_{\beta}\rho_{\beta n}\nu_{n_\beta e}(\vec{u}_{\beta n} - \vec{u}_e) \nonumber \\
&-& \sum_{\beta}\rho_{\beta n}\sum_{\alpha}\nu_{n_\beta i_\alpha}(\vec{u}_{\beta n} - \vec{u}_{\alpha i})
\end{eqnarray}

Finally, the collisional momentum exchange between electrons and the rest of the particles is expressed as:
\begin{eqnarray}  \label{eq:R_e}
\vec{R}_e=&-&\rho_e\sum_{\alpha}\nu_{e i_\alpha}(\vec{u}_e - \vec{u}_{\alpha i}) \nonumber \\
&-&\rho_e\sum_{\beta}\nu_{e n_\beta}(\vec{u}_e - \vec{u}_{\beta n})
\end{eqnarray}
where the summation goes over all ions and neutrals.

The total momentum is conserved in elastic collisions, making $\vec{R}_e+\vec{R}_i+\vec{R}_n=0$.

\section{Definition of variables for two-fluid description}
\label{app:defs2}

We denote the macroscopic parameters for ions ($\mathsc{I}=1$) of the specie $\alpha$ by subscript ``$\alpha i$'', and for neutrals ($\mathsc{I}=0$) by subscript ``$\alpha n$''.  We define the following variables:

\begin{itemize}

\item Total number density of ions and neutrals:
\begin{eqnarray}
n_i =\sum_\alpha n_{\alpha i} = n_e; \,\,\, n_n =\sum_\alpha n_{\alpha n}
\end{eqnarray}
the first equation means global charge neutrality for plasma containing only singly ionized ions.

\item Total mass density of ions and neutrals:
\begin{equation}
\rho_i  = \sum_\alpha n_{\alpha i}m_{\alpha i} ; \,\,\, \rho_n  = \sum_\alpha n_{\alpha n}m_{\alpha n}
\end{equation}
\item Total mass density of charges:
\begin{equation}
\rho_c  =\rho_i+\rho_e= \sum_\alpha n_{\alpha i}m_{\alpha i} + n_em_e
\end{equation}
\item Center of mass velocity of ions and neutrals:
\begin{eqnarray} \label{eq:un-alpha}
\vec{u}_i  = \frac{\sum_\alpha\vec{u}_{\alpha i}\rho_{\alpha i}}{\rho_i}; \,\,\,
\vec{u}_n  = \frac{\sum_\alpha\vec{u}_{\alpha n}\rho_{\alpha n}}{\rho_n}
\end{eqnarray}
\item Center of mass velocity of charges:
\begin{eqnarray} \label{eq:uc-alpha}
\vec{u}_c  = \frac{\sum_\alpha\vec{u}_{\alpha i}\rho_{\alpha i} + \vec{u}_e\rho_e}{\rho_c}=
\frac{\vec{u}_i\rho_i + \vec{u}_e\rho_e}{\rho_c}
\end{eqnarray}
\item Total current density:
\begin{equation} \label{eq:j-alpha}
\vec{J} = e\sum_\alpha n_{\alpha i}\vec{u}_{\alpha i} - en_e\vec{u}_e
\end{equation}
\item Individual drift velocity of species with respect to $\vec{u}_c$ or
    $\vec{u}_n$:
\begin{eqnarray}
\label{eq:wie-alpha}
\vec{w}_{\alpha n}  =   \vec{u}_{\alpha n} - \vec{u}_n; \,
\vec{w}_{\alpha i}  =   \vec{u}_{\alpha i} - \vec{u}_c; \,
\vec{w}_e  =   \vec{u}_e - \vec{u}_c
\end{eqnarray}
Note that the velocity of charges is measured with respect to $\vec{u}_c$ and the velocity of neutrals with respect to $\vec{u}_n$, i.e. the system of reference for charges and neutrals is different.

\item Random velocities of individual species with respect to their means:
\begin{eqnarray}
\label{eq:cie-alpha}
\vec{c}_{\alpha n}  =   \vec{v}_{\alpha n} - \vec{u}_{\alpha n}; \,
\vec{c}_{\alpha i}  =   \vec{v}_{\alpha i} - \vec{u}_{\alpha i}; \,
\vec{c}_e  =   \vec{v}_e - \vec{u}_e
\end{eqnarray}
\noindent where $\vec{v}_{\alpha i}$, $\vec{v}_{\alpha n}$ and
$\vec{v}_e$ are velocities of individual particles.

\item Ion, neutral, and  electron-ion pressure tensors:
\begin{eqnarray}\label{eq:pipe-alpha}
{\bf\hat{p}}_i &=& \sum_\alpha{\bf\hat{p}}_{\alpha i} + \sum_\alpha\rho_{\alpha i}(\vec{w}_{\alpha i}\otimes\vec{w}_{\alpha i}) \\ \nonumber
{\bf\hat{p}}_n &=& \sum_\alpha{\bf\hat{p}}_{\alpha n} + \sum_\alpha\rho_{\alpha n}(\vec{w}_{\alpha n}\otimes\vec{w}_{\alpha n}) \\ \nonumber
{\bf\hat{p}}_{ie} & = & {\bf\hat{p}}_i + {\bf\hat{p}}_e+ \rho_e(\vec{w}_e\otimes\vec{w}_e)
\end{eqnarray}
\item Ion, neutral and electron-ion  pressure scalars:
\begin{eqnarray}
p_i&=&\sum_\alpha p_{\alpha i} + \frac{1}{3}\sum_\alpha{\rho_{\alpha i} w_{\alpha i}^2} \nonumber \\
p_n &=& \sum_\alpha p_{\alpha n} + \frac{1}{3}\sum_\alpha{\rho_{\alpha n} w_{\alpha n}^2} \nonumber \\
p_{ie}&=&p_i+ p_e + \frac{1}{3}{\rho_ew_e^2}
\end{eqnarray}
\item Ion, neutral, and electron-neutral heat flow vectors:
\begin{eqnarray}  \label{eq:qie-alpha}
\vec{q}_i &=& \sum_\alpha \left( \vec{q}_{\alpha i} + {\bf\hat{p}}_{\alpha i}  \vec{w}_{\alpha i}  +  \frac{3}{2} p_{\alpha i}\vec{w}_{\alpha i} + \frac{1}{2}\rho_{\alpha i}w_{\alpha i}^2\vec{w}_{\alpha i} \right) \nonumber \\
\vec{q}_n &=&\sum_\alpha \left( \vec{q}_{\alpha n} + {\bf\hat{p}}_{\alpha n}  \vec{w}_{\alpha n}  + \frac{3}{2}p_{\alpha n}\vec{w}_{\alpha n} + \frac{1}{2}\rho_{\alpha n}w_{\alpha n}^2\vec{w}_{\alpha n} \right ) \nonumber \\
\vec{q}_{ie} &=& \vec{q}_i+ \vec{q}_e + {\bf\hat{p}}_e  \vec{w}_e + \frac{3}{2} p_e\vec{w}_e+ \frac{1}{2}\rho_e w_e^2\vec{w}_e
\end{eqnarray}
\item Ion, neutral, and electron-neutral heat flow vectors corrected for potential ionization-recombination energy flux:
\begin{eqnarray}  \label{eq:qie-alpha-prime}
\vec{q}^{\prime}_i &=& \vec{q}_i + \sum_\alpha \chi_{\alpha i} \vec{w}_{\alpha i}\nonumber \\
\vec{q}^{\prime}_n &=& \vec{q}_n + \sum_\alpha \chi_{\alpha n} \vec{w}_{\alpha n}\nonumber \\
\vec{q}^{\prime}_{ie}&=& \vec{q}_{ie} + \sum_\alpha \chi_{\alpha i} \vec{w}_{\alpha i}
\end{eqnarray}

\end{itemize}

\section{Definition of variables for single-fluid description}
\label{app:defs1}

\noindent Below the summation is be done over $N$ neutrals, $N$ ions and one electron component, indexed by $\alpha=1...2N+1$. We make use of the definitions of the total ion and neutral densities and velocities from Appendix \ref{app:defs2}.

\begin{itemize}

\item Total mass density:
\begin{equation}
\label{eq:totalrho} \rho  = \sum_{\alpha=1}^{2N+1}n_{\alpha}m_{\alpha}= \rho_n + \rho_i + \rho_e
\end{equation}
\item Center of mass velocity.
\begin{equation}
\label{eq:totalu} \vec{u}  = \frac{\sum_{\alpha=1}^{2N+1}(\rho_{\alpha}\vec{u}_{\alpha})}{\rho} = \frac{\rho_n\vec{u}_n + \rho_i\vec{u}_i + \rho_e\vec{u}_e}{\rho}
\end{equation}
\item Drift velocity. The drift velocity of each specie is taken with respect to the average center of mass velocity $\vec{u}$.
\begin{eqnarray}
\vec{w}_{\alpha} & = & \vec{u}_{\alpha} - \vec{u} \label{eq:w1}
\end{eqnarray}
where $\alpha=1...2N+1$. Note, that $\sum_{\alpha}\rho_{\alpha}\vec{w}_{\alpha} = 0$
\item Total current density:
\begin{eqnarray} \label{eq:totalj}
\vec{J}& = &e\sum_{\alpha=1}^Nn_{\alpha}\vec{u}_{\alpha} - en_e\vec{u}_e
\end{eqnarray}
where the summation goes only over $N$ ions.
\item Total kinetic pressure tensor:
\begin{eqnarray}
{\bf\hat{p}} =  \sum_{\alpha=1}^{2N+1}\hat{p}_{\alpha} + \sum_{\alpha=1}^{2N+1}\rho_{\alpha}\vec{w}_{\alpha}\otimes\vec{w}_{\alpha}
\label{eq:totalpkin1}
\end{eqnarray}
\item Total pressure scalar:
\begin{equation}
p = \sum_{\alpha=1}^{2N+1}p_{\alpha} + \frac{1}{3}\sum_{\alpha=1}^{2N+1}\rho_{\alpha}w_{\alpha}^2
\label{eq:totalpscalar1}
\end{equation}
\item Total heat flow vector:
\begin{eqnarray} 
\label{eq:totalq1} 
\vec{q} =  \sum_{\alpha=1}^{2N+1}\left(\vec{q}_{\alpha} + {\bf\hat{p}}_{\alpha}  \vec{w}_{\alpha} + 
\frac{3}{2}p_{\alpha}\vec{w}_{\alpha} +\frac{1}{2}\rho_{\alpha}w_{\alpha}^2\vec{w}_{\alpha} \right)
\end{eqnarray}
\item Total heat flow vector corrected for potential ionization-recombination energy flux:
\begin{eqnarray}   \label{eq:q-prime}
\vec{q}^{\prime} =\vec{q}  + \sum_{\alpha=1}^{2N}\chi_{\alpha}\vec{w}_\alpha
\end{eqnarray}
with $\chi_\alpha$ given by Eq. \ref{eq:chi-ai} without explicit indication of ions and neutrals.

\end{itemize}

\providecommand{\noopsort}[1]{}\providecommand{\singleletter}[1]{#1}%

\end{document}